\documentclass[11pt]{article}
\usepackage{graphicx}
\usepackage{latexsym,graphicx}
\usepackage{float}
\usepackage{amsmath}
\usepackage{amsfonts}
\usepackage{amssymb}
\usepackage{epstopdf}
\usepackage{color}
\DeclareGraphicsExtensions{.eps}
\usepackage[left=2.5cm,top=2.5cm,right=2.5cm,bottom=2.5cm]{geometry}

\catcode `\@=11
\catcode `\@=12
\begin{document}
    	\begin{center}
    		\large{{\bf Reconstruction of an Observationally Constrained $f(R, T)$ gravity model}} \\
    		\vspace{5mm}
    		\normalsize{Anirudh Pradhan$^1$, Gopikant Goswami$^2$, Aroonkumar  Beesham$^{3,4}$}\\
    		\vspace{5mm}
    		\normalsize{$^{1}$Centre for Cosmology, Astrophysics and Space Science (CCASS), GLA University, Mathura-281 406, Uttar Pradesh, India}\\
    		\vspace{5mm}
    		\normalsize{$^{2}$Department of Mathematics, Netaji Subhas University of Technology, Delhi, India}\\
    		\vspace{5mm}
    		\normalsize{$^{3}$Department of Mathematical Sciences, University of Zululand Private Bag X1001 Kwa-Dlangezwa 3886 South Africa}\\
    		\vspace{5mm}
    		\normalsize{$^{4}$Faculty of Natural Sciences, Mangosuthu University of Technology, P O Box 12363, Jacobs 4052, South Africa}\\
    		\vspace{2mm}
    		$^1$E-mail: pradhan.anirudh@gmail.com \\
    		\vspace{2mm}
    		$^2$E-mail: gk.goswami9@gmail.com \\
    		\vspace{2mm}
    		$^{3,4}$E-mail: abeesham@yahoo.com \\
    	\end{center}
    	\begin{abstract}
    		In this paper, an attempt is made to construct a  Friedmann-Lemaitre-Robertson-Walker model in $f(R,T)$  gravity with a perfect fluid that yields acceleration at late times. We take  $f(R,T)$ as $R$ + $8\pi \mu T$. As in the  $\Lambda$CDM model, we take the matter to consist of two components, viz., $\Omega_m$ and $\Omega_{\mu}$ such that $\Omega_m$ + $\Omega_{\mu}$=1. The parameter $\Omega_m$ is the matter density (baryons + dark matter), and $\Omega_{\mu}$ is the density associated with the Ricci scalar $R$ and the  trace $T$ of the energy momentum tensor, which we shall call dominant matter.  We find that at present $\Omega_{\mu}$ is dominant over $\Omega_m$, and that the two are  in the ratio 3:1 to 3:2 according to the three data sets:  (i) 77 Hubble OHD data set (ii) 580 SNIa supernova distance modulus data set and (iii) 66 pantheon SNIa data which include high red shift data in the range $0\leq z\leq 2.36$. We have also calculated the pressures and densities associated with the two matter densities, viz., $p_{\mu}$, $\rho_{\mu}$, $p_m$ and $\rho_m$, respectively. It is also found that  at present, $\rho_{\mu}$ is greater than  $\rho_m$. The negative dominant matter  pressure $p_{\mu}$ creates acceleration in the universe. Our deceleration and snap  parameters show a change from negative to positive, whereas the jerk parameter is always positive. This means that the universe is at present accelerating and in the past it was decelerating. State finder diagnostics indicate that our model is at present a dark energy quintessence model. The various other physical and geometric properties of the model are also discussed. 
    		\end{abstract}
    		{\bf Keywords}: $ f(R,T)$ theory; FLRW metric; Observational parameters; Transit universe; Observational constraints \\ 
    	
    	Mathematics Subject Classification 2020: 83D05, 83F05, 83C15  \\ 
    	
    	
\section{Introduction:} Cosmology is the study of the large scale structure and evolution of the universe.  It began seriously with the simple Einstein static universe \cite{ref1}. It then got a drastic change due to Hubble \cite{ref2}, who started the concept of an expanding universe. This was later  mathematically formulated into  the FLRW (Friedmann-Lemaitre-Robertson-Walker) spacetime \cite{ref3,ref4,ref5,ref6,ref7}. The universe since its inception, passed through very exotic events such as inflation \cite{ref8,ref9}  at the beginning, then the discovery of the CMB (cosmic microwave background radiation) \cite{ref10}, and then the late time acceleration\cite{ref11}$-$\cite{ref26}. Dark energy, which is believed to cause this acceleration is most often represented by the very simple concordance $\Lambda$CDM model. This model fits well on observational grounds \cite{ref27}$-$\cite{ref29},  despite certain weaknesses \cite{ref30} that it suffers with fine tuning and the cosmic coincidence problems. To solve these, scalar field dominated tracker field  quintessence and phantom  dark energy models were proposed \cite{ref31}$-$\cite{ref34}.  Later on, parameterizations for the scalar $\phi$ field were suggested, and some interesting  cosmological models were developed \cite{ref35} $-$ \cite{ref42}. This work has been done under the principle of general relativity (GR) and accordingly, the GR field equations were used.
	
With this, a spate of work was begun in which attempts were made to get an accelerating universe by modifying Einstein's field equations. It has been suggested that the Ricci scalar $R$ be replaced by an arbitrary function $f(R)$ of $R$ in the Einstein-Hilbert action,  and that the new theory and its field equations be called $f(R)$ theory \cite{ref43} $-$ \cite{ref56}. The idea is that a non linear Ricci scalar may be helpful in developing negative pressure in the universe, producing acceleration. Seeing the complexity of solutions due to non linear Ricci scalar \cite{ref57}, more alternatives were proposed. In one of the options, $f(R)$ was replaced by an arbitrary function $f(R,T)$ of $R$ and $T$ where $T$ is the trace of the energy momentum tensor. This theory is called $f(R,T)$ gravity \cite{ref58} $-$ \cite{ref70}. The authors of the theory \cite{ref58}  have dealt three options for the specific functional form of $f(R,T$), especially for cosmological interpretations. They are $R + 2f(T), f_{1}(R) + f_{2}(T) ~ \text{and} 	~ f_{1}(R) + f_{2}(R)f_{3}(T)$.
The purpose of this work is to model a universe in  $f(R,T)$ gravity which meets observational constraints \cite{ref20}$-$\cite{ref22}.
For this, we consider the first simple alternative of $f(R,T)$, i.e., $f(R,T) = R + \lambda$ T, taking  $\lambda$ as an arbitrary constant. We consider an FLRW spacetime with perfect fluid.  Our three model parameters, viz., the Hubble, deceleration and equation of state, are estimated with the help of three data sets: (i) the 77 Hubble OHD data set \cite{ref71}$-$\cite{ref86} (ii) the 580 SNIa supernova distance modulus data set \cite{ref17} and (iii) the  66 pantheon SNIa data set,  which includes the high red shift data in the range $0\leq z\leq 2.36$ \cite{ref87}$-$\cite{ref89}. We solve the $f(R,T)$ field equations by making the simplest possible parametrization of the equation of state parameter as given by Gong and Zhang \cite{ref35}.\\
 At this junction,  it is desirable to describe the new results/important points presented in this paper which are different from the past studies. In the past there had been some very interesting reconstruction and review works in the modified theories of gravity \cite{ref69}$-$ \cite{ref72}. The learned authors also noticed certain finite time singularities and rips in certain models. However as an important and noticeable feature in our model, we have  developed two energy parameters $\Omega_m$ and $\Omega_{\mu}$ and found  that $\Omega_m$ + $\Omega_{\mu}$=1. The parameter $\Omega_m$ is associated with the matter, whereas $\Omega_{\mu}$ is associated with $f(R,T)$ gravity.  We have statistically estimated using high red shift data set in the range $0\leq z\leq 2.36$  that at present $\Omega_{\mu}$ is dominant,  and that the two energy densities are approximately in the ratio 3:1 to 3:2. Moreover we have also performed a state finder diagnostic of our model  and found  that our model is at present in quintessence. We have presented 35 plots and tables to describe the findings of our work more effectively.\\
	 
The following is a short description of the work, broken down into sections:
In section $2$, the $f(R,T)$ field equation for a perfect fluid filled FLRW space time are described. Section $3$ is the architect of the rest of the part of the paper. In this section, we get expressions for the  Hubble and deceleration parameters as functions of red shift.  In sections $4$, $5$ and $6$, the model parameters  have been estimated on the basis of the three data sets. In sections 7, we have presented the error bar plots, confidence regions and likelihood plots for the Hubble, distance modulus and apparent magnitude. The purpose here is  to show the proximity of observational and theoretical results. The  jerk and snap parameters are discussed in section 8, and it is found that our model is at present a dark energy quintessence model. 
In section $9$, we have obtained expressions for the densities and pressures. It is found that at present $\mu$ density ($\rho_{\mu}$) is dominant over baryon density and they are nearly in the ratio  1:3 or 2:3 as per different observed  data sets. The last section $10$ is devoted to  the concluding remarks. 
	 
	  
\section{f(R,T) Field Equations for Perfect Fluid Filled FLRW Space Time:} We refer to \cite{ref58} for  the field equations of f(R,T) gravity:
\begin{equation} {\label{1}}
		R_{ij}-\frac{1}{2} R g_{ij} = \frac{8 \pi G T_{ij}}{f^R(R,T)}+ \frac{1}{f^R (R,T)} \bigg(\frac{1}{2} g_{ij} (f(R,T)-R f^R (R,T)) - (g_{ij} \Box - \nabla_i \nabla_j) f^R (R,T) + f^T (R,T) (T_{ij} +p g_{ij}) \bigg),
\end{equation}
	where we take the energy momentum tensor $T_{ij}$ as that of perfect fluid:
	\begin{equation}\label{2}
		T_{ij}= (\rho + p) u_i u_j - p g_{ij}.
	\end{equation}
  	The other symbols in the field equations (\ref{1}) have their usual meanings.   	   The Einstein Hilbert action is:
  	  \begin{equation}{\label{3}}
  	  	S= \int(\frac{1}{16\pi G}( R+2\lambda) + L_m)\sqrt{-g} dx^4,
  	  \end{equation}
     where $L_m$ is the  matter Lagrangian. The field equations (\ref{1}) are obtained replacing the Ricci scalar R by a arbitrary function $f(R,T)$ of $R$ and trace $T$ of energy momentum tensor $T_{ij}$. Accordingly, the $f(R,T)$ action is:
     \begin{equation}{\label{4}}
     	S= \int{\left(\frac{1}{16\pi G} f(R,T)+L_m \right)\sqrt{-g}} dx^4.
     \end{equation}
    	The FLRW metric is:
	\begin{equation}{\label{5}}
		ds^{2} = dt^{2} - a^{2}(t) (dx^{2} + dy^{2} + dz^{2}).
		\end{equation}
	We solve Eqs. (\ref{1}) and (\ref{2}) for this metric and get:
	\begin{equation}{\label{6}}
		2 \dot{H} + 3 H^2 = -(8 \pi + 3 \lambda) p + \lambda \rho
	\end{equation}
	and
	\begin{equation}{\label{7}}
		3 H^2 = (8 \pi + 3 \lambda ) \rho - \lambda p,
	\end{equation}
	where $H=\frac{\dot{a}}{a}$ is the Hubble parameter.
	
	\section{Deceleration and Hubble Parameters:}
	We  assume $ \lambda = 8 \pi \mu $. Then the field equations  (\ref{6}) and (\ref{7}) are simplified as:
	
	\begin{equation}{\label{8}}
	H^2 (1-2 q)=-8 \pi  ( p +  \mu  (3 p-\rho ))
	\end{equation}
	and
	\begin{equation}{\label{9}}
	3 H^2=8 \pi  ( \rho +  \mu  (3 \rho -p)),
	\end{equation}
    where $q=-\frac{\ddot{a}}{a H^2}$ is deceleration parameter and $ \dot{H} = - (q+1)H^2 $.
    It is clear from the field equations  (\ref{8}) and (\ref{9}) that for $\mu =0$, these equations reduce to those of general relativity. So the additional terms containing  $\mu$  are there due to $f(R,T)$ gravity. We require $q\leq 0$ (acceleration) at present due to the presence of these terms. We let:
    \begin{equation}\label{10}
    p_\mu \equiv \mu  (3 p-\rho ), ~ \rho_\mu \equiv \mu  (3 \rho -p).
    \end{equation}
    and call these the dominant pressure and dominant energy density, respectively, for reasons which will become clear later. These may be regarded as the contribution of  $f(R,T)$ gravity to the pressure and density. We get the  energy parameters  $ \Omega _m$ and $\Omega _{\mu}$ and the equations of state parameter $\omega$ and  $\omega_\mu$ as follows:
    \begin{equation}\label{11}
    \Omega _m=\frac{8\pi\rho}{3H^2}, ~  \Omega _{\mu}=\frac{8\pi \rho_\mu}{3H^2}, ~ \omega_m =\frac{p}{\rho},~ \omega_\mu =\frac{p_\mu}{\rho_\mu}.
    \end{equation}
	Here suffixes $m$ and $\mu$ stands for the GR effect, and $f(R,T)$ effect, respectively. We may also interpret the  parameters with the $\mu$  suffix as terms arising due to the curvature dominance of $f(R,T)$ gravity.

	From Eqs. (\ref{8}) $-$ (\ref{11}), we get the following:
	\begin{equation}\label{12}
		H^2 (1-2 q)=-8 \pi  ( p +  p_\mu), ~~ 	3 H^2=8 \pi  ( \rho + \rho_\mu ),~~ \Omega _m +	 \Omega _{\mu} =1
	\end{equation}
	and
	\begin{equation}\label{13}
	\omega _{\mu }=\frac{3 \omega _m-1}{3-\omega _m} ~~~~ \omega _m = \frac{3 \omega _{\mu }+1}{\omega _{\mu }+3}.	
	\end{equation}
    Eqs. (\ref{12}) and (\ref{13})  give us the following expression for  $q$ :
    	\begin{equation}\label{14}
    	2 q = 1+ \frac{3 (8 \mu +3) \omega _{\text{$\mu $}}+3}{8 \mu +\omega _{\text{$\mu $}}+3}.	
    \end{equation}
     	From this, we get the present value of $\mu$ as: 
	\begin{equation}\label{15}
		\mu_0 = \frac{5 \omega _{\text{$\mu $0}} - q_0 \left(\omega _{\text{$\mu $0}}\right)-3 q_0+3}{4 \left(-3 \omega _{\text{$\mu $0}}+2 q_0-1\right)}
	\end{equation}	
	   We observe that there are  two  equations  and four unknowns $H$, $q$, $p$ and $\rho$. Hence, one cannot solve these equation in general. However, to get an explicit solution of the above equations, we have to assume at least one reasonable relation among the
	variables. For this, we consider the simplest parametrization of the equation of state parameter $\omega_{\mu} $ as given by Gong and Zhang \cite{ref35}:
	\begin{equation}\label{16}
	\omega_\mu = \frac{\omega_{\mu 0}}{(1+z)},	
	\end{equation}
     where $z$ is the red shift and $\omega_{\mu 0}$ is the present value of the equation of state parameter $\omega_{\mu}$ . With this we can now solve for $q$ in term of $z$ as follows:
     \begin{equation}\label{17}
     	q =\frac{q_0 \left(1.5 \omega _{\text{$\mu $0}}^2+0.5 z \omega _{\text{$\mu $0}}-1.5 z-1.5\right)+2 z \omega _{\text{$\mu $0}}}{1.5 \omega _{\text{$\mu $0}}^2+\left(q_0-0.5\right) z \omega _{\text{$\mu $0}}-1.5 z-1.5}
     \end{equation}
 Then H may be obtained from q by solving the following differential equation:
 \begin{equation}\label{18}
 	H_z(1+z)=(q+1)H	
 \end{equation}
 We obtain the expression for $H$ as follows:
 \begin{equation}\label{19}
 	H = H_{0}~ \alpha_{1}~ \alpha_{2}~ exp{\left[-\frac{\alpha_3 \alpha_4}{\alpha_5 }+\frac{\alpha_6 \alpha_7}{\alpha_8}\right]},
 \end{equation}
where, $H_0$ is the present value of the Hubble parameter and:
\vspace{.5cm}
	
$\begin{array}{rrrrr}

\alpha_1&=&\left(1.5\, -1.5 \omega _{\text{$\mu $0}}^2\right){}^{\frac{q_0 \left(0.75\, -0.75 \omega _{\text{$\mu $0}}\right)-0.75 \omega _{\text{$\mu $0}}+0.75}{\left(q_0-0.5\right) \omega _{\text{$\mu $0}}-1.5}}\hspace{35cm}\\
\\
\alpha_2 &=&\left(z \omega _{\text{$\mu $0}} \left(q_0 (-1. z-1.)+0.5 z+0.5\right)+(-1.5 z-1.5) \omega _{\text{$\mu $0}}^2+z (1.5 z+3)+1.5\right){}^{\frac{-0.75 \omega _{\text{$\mu $0}}-0.75 q_0 \omega _{\text{$\mu $0}}+0.75 q_0+0.75}{0.5 \omega _{\text{$\mu $0}}-1. q_0 \omega _{\text{$\mu $0}}+1.5}}\hspace{25cm}\\
\\
 \alpha_3&=&\left(q_0 \left(3.75\, -1.5 q_0\right)+5.25\right) \omega _{\text{$\mu $0}}+\left(-0.750 q_0^2-3.75\right) \omega _{\text{$\mu $0}}^3-1.5 \left( q_0+1\right){}^2 \omega _{\text{$\mu $0}}^2+ 8.88^{-16} (q_0+1)\hspace{25cm}\\
 \\
 \alpha_4&=&\tan ^{-1}\left(\frac{\left(0.5\, -1. q_0\right) \omega _{\text{$\mu $0}}-1.5 \omega _{\text{$\mu $0}}^2+3}{\sqrt{\omega _{\text{$\mu $0}} \left(\omega _{\text{$\mu $0}} \left(\left(-2.25 \omega _{\text{$\mu $0}}-1.5\right) \omega _{\text{$\mu $0}}-0.25\right)+q_0^2 \left(-\omega _{\text{$\mu $0}}\right)+q_0 \left(\omega _{\text{$\mu $0}} \left(3. \omega _{\text{$\mu $0}}+1.\right)-8.88^{-16}\right)+4.44^{-16}\right)+1.78^{-15}}}\right)\hspace{25cm}\\
 \\
 \alpha_5&=& c \sqrt{\omega _{\text{$\mu $0}} \left(\omega _{\text{$\mu $0}} \left(\left(-2.25 \omega _{\text{$\mu $0}}-1.5\right) \omega _{\text{$\mu $0}}-0.25\right)-q_0^2 \omega _{\text{$\mu $0}}+q_0 \left(\omega _{\text{$\mu $0}} \left(3. \omega _{\text{$\mu $0}}+1.\right)-8.88^{-16}\right)+4.44^{-16}\right)+1.77^{-15}}\hspace{25cm}\\
  c&=&\left(0.5 \omega _{\text{$\mu $0}}-q_0 \omega _{\text{$\mu $0}}+1.5\right)\hspace{35cm}\\
  \\
  \alpha_6 &=& 3. \left(-1.25 \omega _{\text{$\mu $0}}^3-0.5 \omega _{\text{$\mu $0}}^2+1.75 \omega _{\text{$\mu $0}}+q_0^2 \left(\omega _{\text{$\mu $0}}^2-0.5 \omega _{\text{$\mu $0}}-0.5\right) \omega _{\text{$\mu $0}}\right)\hspace{30cm}\\
 && + 3\left(q_0 \left(-0.25 \omega _{\text{$\mu $0}}^3-\omega _{\text{$\mu $0}}^2+1.25 \omega _{\text{$\mu $0}}+2.96^{-16}\right)+2.96^{-16}\right)\hspace{30cm}\\
  \\
  \alpha_7&=&\tan ^{-1}\left(\frac{-1.5 \omega _{\text{$\mu $0}}^2+\omega _{\text{$\mu $0}} \left(q_0 (-2. z-1.)+1. z+0.5\right)+3. z+3.}{\sqrt{-2.25 \omega _{\text{$\mu $0}}^4+\left(3. q_0-1.5\right) \omega _{\text{$\mu $0}}^3+\left(-1. q_0^2+1. q_0-0.25\right) \omega _{\text{$\mu $0}}^2+\left(4.44^{-16}-8.88^{-16} q_0\right) \omega _{\text{$\mu $0}}+1.77^{-15}}}\right)\hspace{25cm}\\
  \\
 \alpha_8&=&\left(0.5 \omega _{\text{$\mu $0}}-q_0 \omega _{\text{$\mu $0}}+1.5\right)\hspace{35cm}\\
  &&\sqrt{-2.25 \omega _{\text{$\mu $0}}^4+\left(3 q_0-1.5\right) \omega _{\text{$\mu $0}}^3+\left(-q_0^2+q_0-0.25\right) \omega _{\text{$\mu $0}}^2+\left(4.44^{-16}-8.88^{-16} q_0\right) \omega _{\text{$\mu $0}}+1.77^{-15}}\hspace{25cm}
  \end{array}$

\section{Estimation of $H_0$, $q_0$ and $\omega_{\mu 0}$ from A Observed Hubble Data Set:}
In this section we will estimate the present values of $H_0$, $ q_0$ and $ \omega_{\mu 0}$ statistically from  a data set of $77$ observed values of H at different red shifts using (a) The cosmic chronometric method, (b) the  BAO signal in galaxy distribution and (c) the BAO signal in the Ly$\alpha$ forest distribution alone or cross-correlated with QSOs \cite{ref71}$-$\cite{ref86}. We compare the data set values from those obtained theoretically from Eq. (\ref{19}) by forming the following chi square function of   $H_0$, $q_0$ and $\omega_{\mu 0}$. 
\begin{equation}\label{20}
	\chi^{2}( H_0, q_0, \omega{\mu 0}) =\frac{1}{76} \sum\limits_{i=1}^{77}\frac{[Hth(z_{i}, H_0, q_0, \omega{\mu 0}) - H_{ob}(z_{i})]^{2}}{\sigma {(z_{i})}^{2}},
\end{equation}
These parameters are estimated by getting the minimum value of chi squared for the values of the parameters  taken in the range ( $ H_0$,65$-$75),  ($q_0$, -0.60 $-$ -0.40) and ($\omega_{\mu 0}$, -0.90 $-$ -0.70) . It is found that 
$$\left\{H_0= 69.2163,q_0 = -0.551698,\omega _{\text{$\mu $0}} =-0.759826\right\}$$ 
for minimum chi squire $\chi2$ =0.685263 which is a a good fit. The fit will be more clear from the following error bar and livelihood plots in figure $1$.

Figures $1(a)$ and $1(b)$ describe the growth of $H$ and  $H/(1+z) = \dot{a}/a_0$ over  red shift $ z$,  respectively. H is an increasing function of z.  These figures also show the closeness of the theoretical graph to the observed values.   Figures $1(c)$, $1(d)$ and $1(e)$ are likelihood probability curves for  $H_0$,  $ q_0$ and  $ \omega_{\mu 0}$. Estimated values  on the basis of the 77 OHD data set are  $\left\{H_0= 69.2163, q_0 = -0.551698, \omega _{\text{$\mu $0}} =-0.759826\right\}$.

\begin{figure}[H]
	(a)	\includegraphics[width=8cm,height=7cm,angle=0]{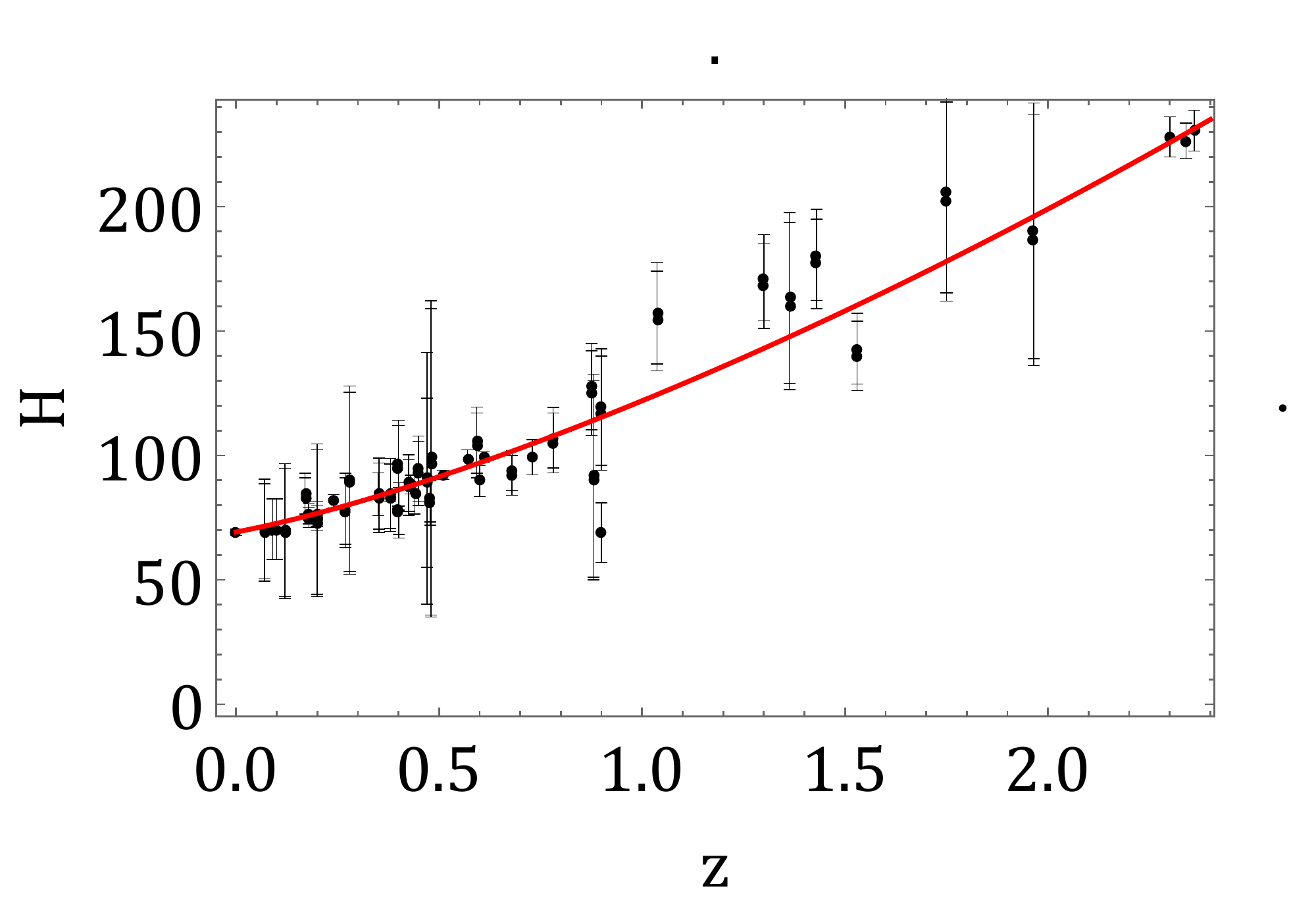}
	(b) \includegraphics[width=8cm,height=7cm,angle=0]{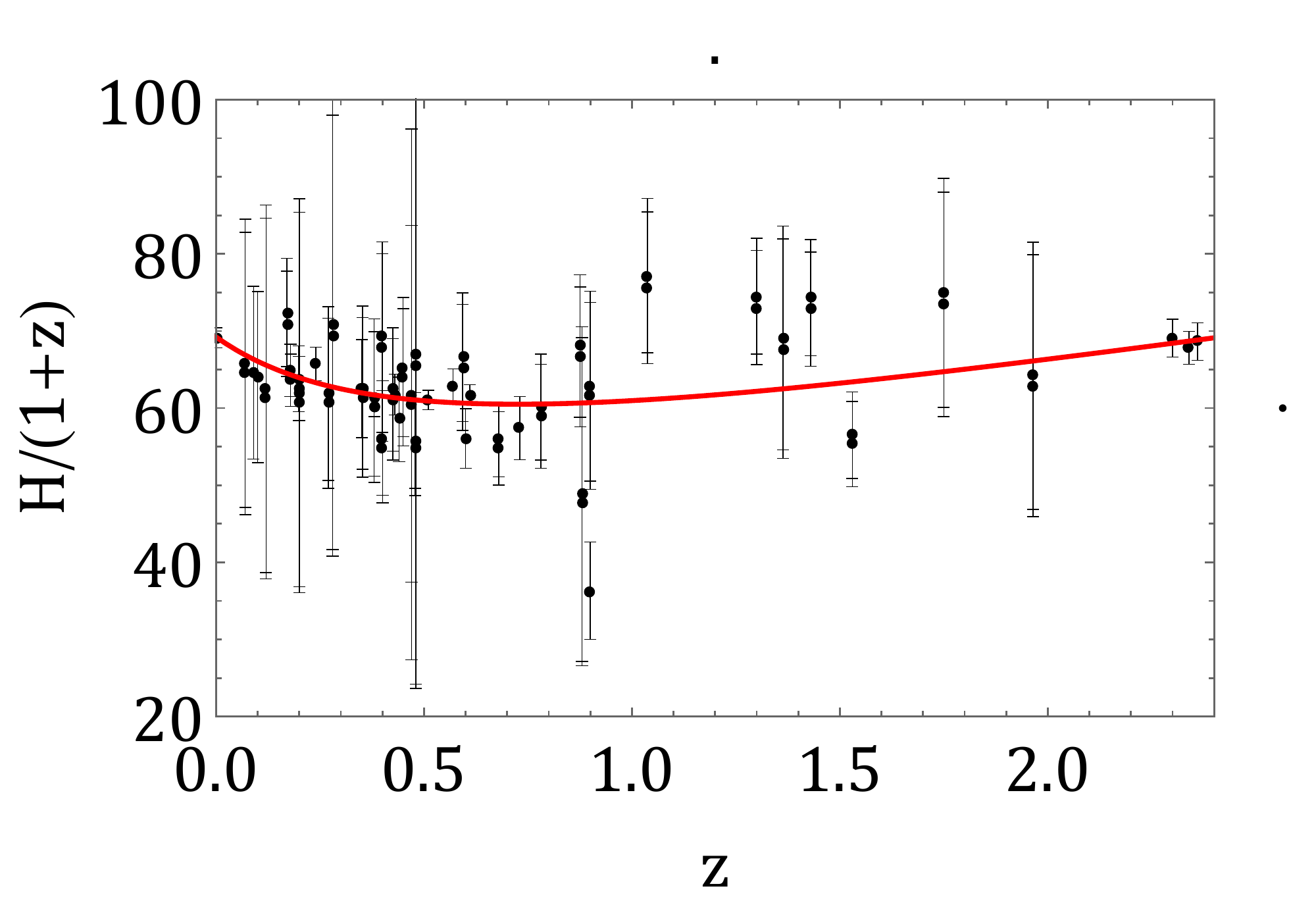}
	(c) \includegraphics[width=8cm,height=7cm,angle=0]{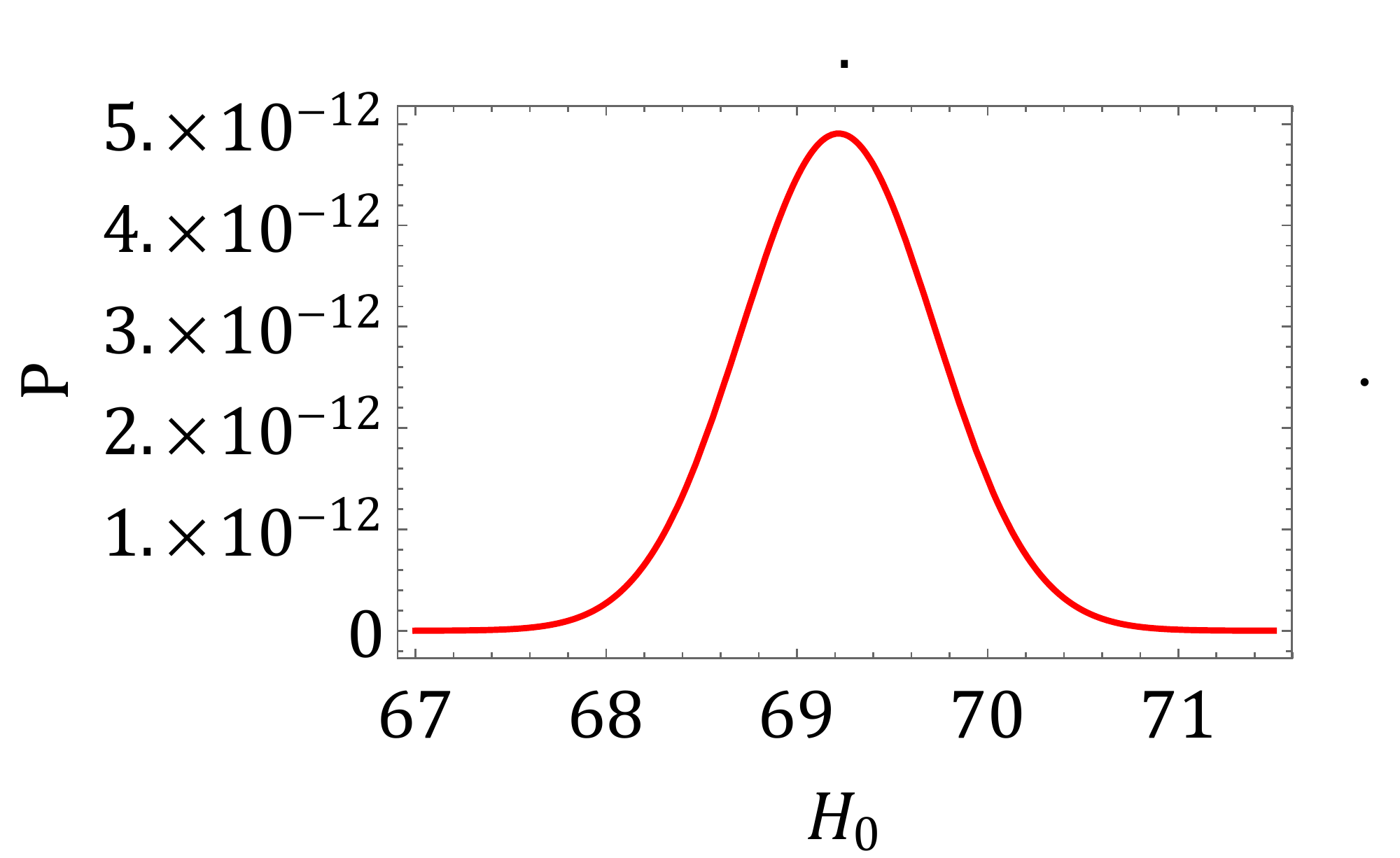}
	(d)	\includegraphics[width=8cm,height=7cm,angle=0]{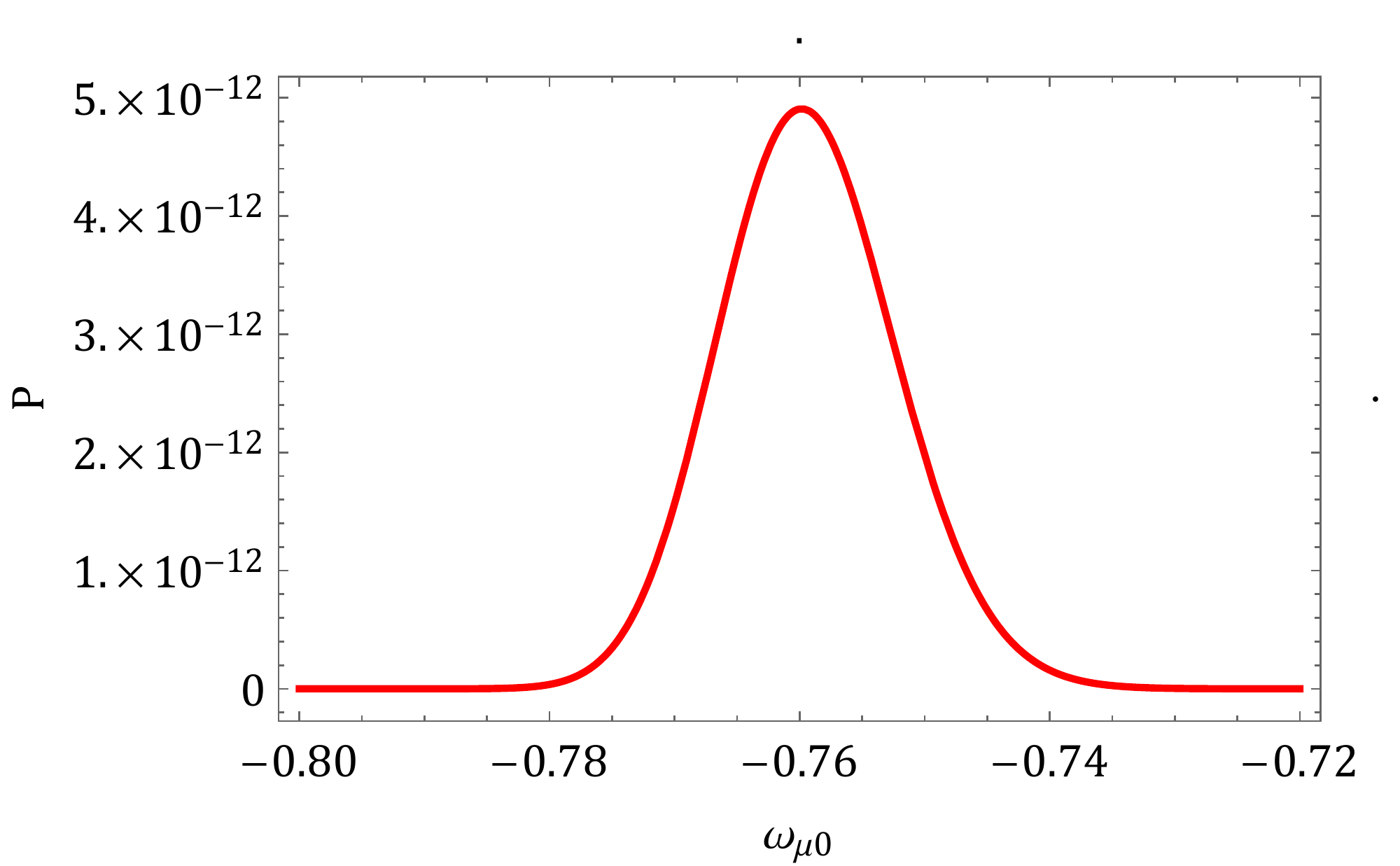} \\
	(e) \includegraphics[width=8cm,height=7cm,angle=0]{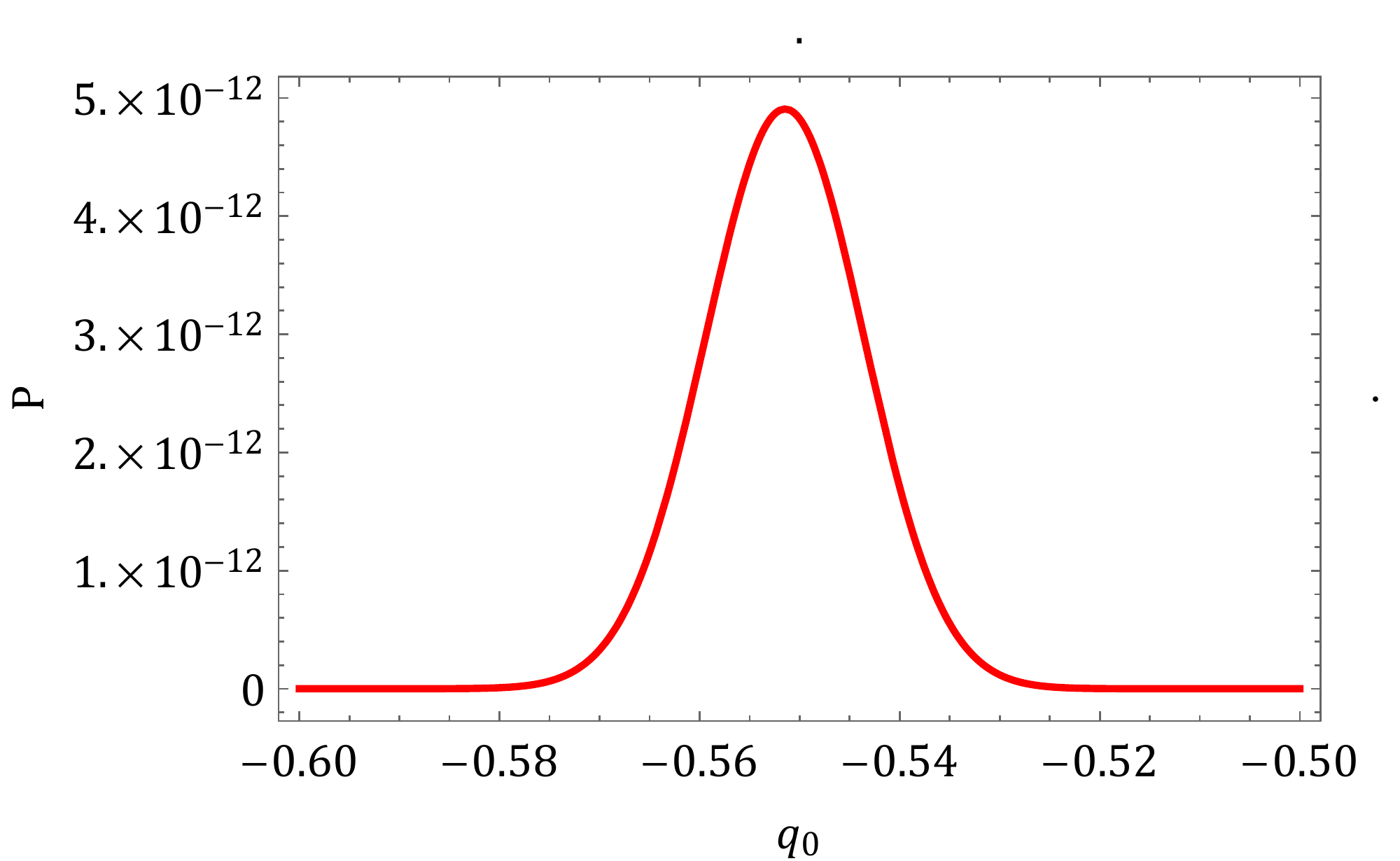}
	\caption{Figures (a) and (b) are the error bar plots  for  Hubble parameter $H$ and  expansion rate $H/(1+z) = \dot{a}/a_0$ over red shift $ z$,  respectively. Figures (c), (d) and (e)  are likelihood probability curves for  $H_0$, $ q_0$ and  $ \omega_{\mu 0}$.}
\end{figure}
 Figures $2(a)$, $2(b)$, and $2(c)$ are the 1$\sigma$, 2$\sigma$ and 3$\sigma$ confidence region plots for the pair of parameters ($H_0$, $q_0$ ), ($H_0$, $\omega_{\mu 0}$) and ($q_0$, $\omega_{\mu 0}$). The estimated value points $(H_0= 69.2163,q_0 = -0.551698)$,$(H_0= 69.2163, \omega _{\text{$\mu $0}} =-0.759826)$ and $(q_0 = -0.551698,\omega _{\text{$\mu $0}} =-0.759826)$ are red spotted. These are just there to show that our estimated values are within the statistically specified regions. 
\begin{figure}[H]
	(a)	\includegraphics[width=9cm,height=8cm,angle=0]{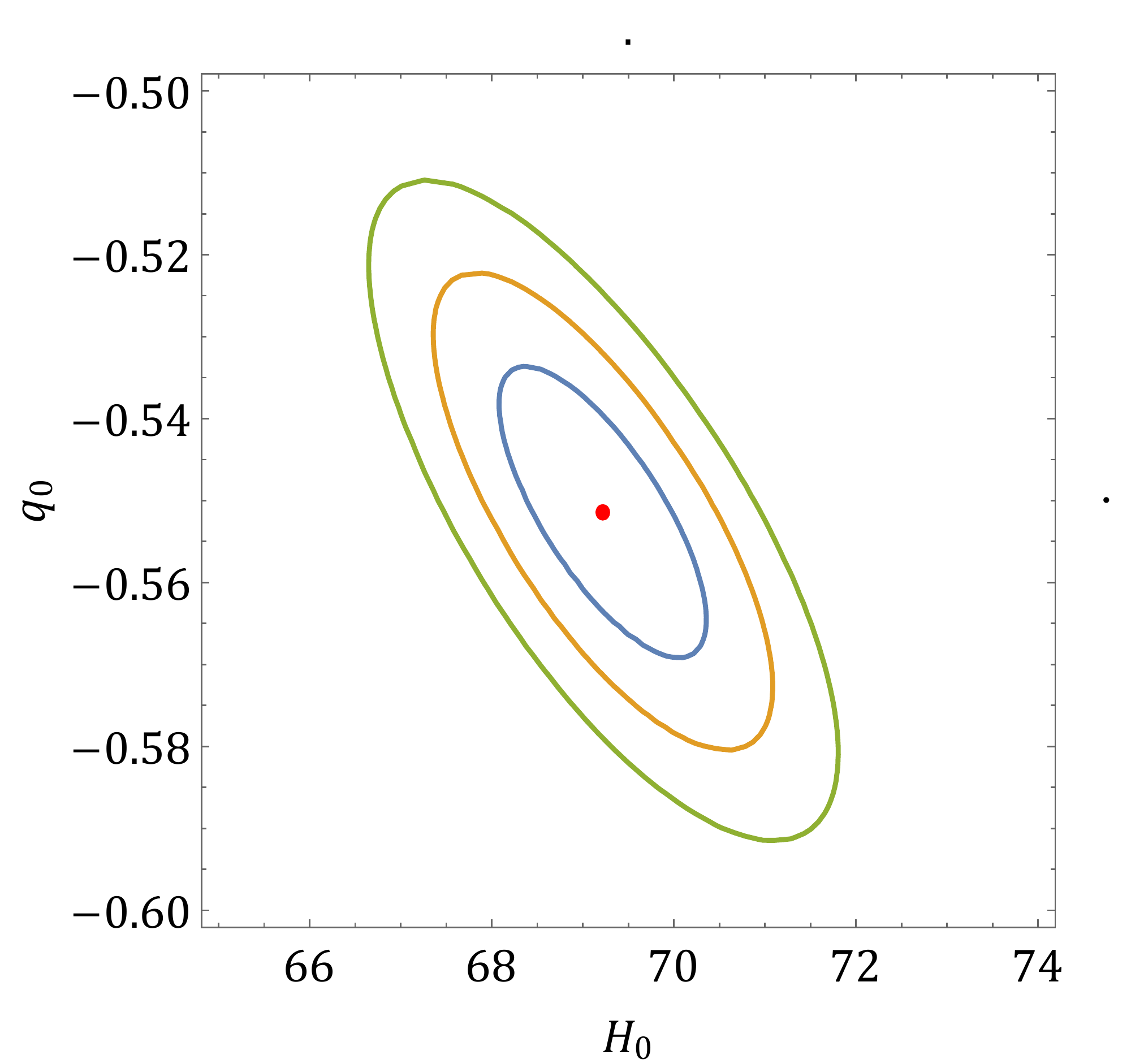}
	(b) \includegraphics[width=9cm,height=8cm,angle=0]{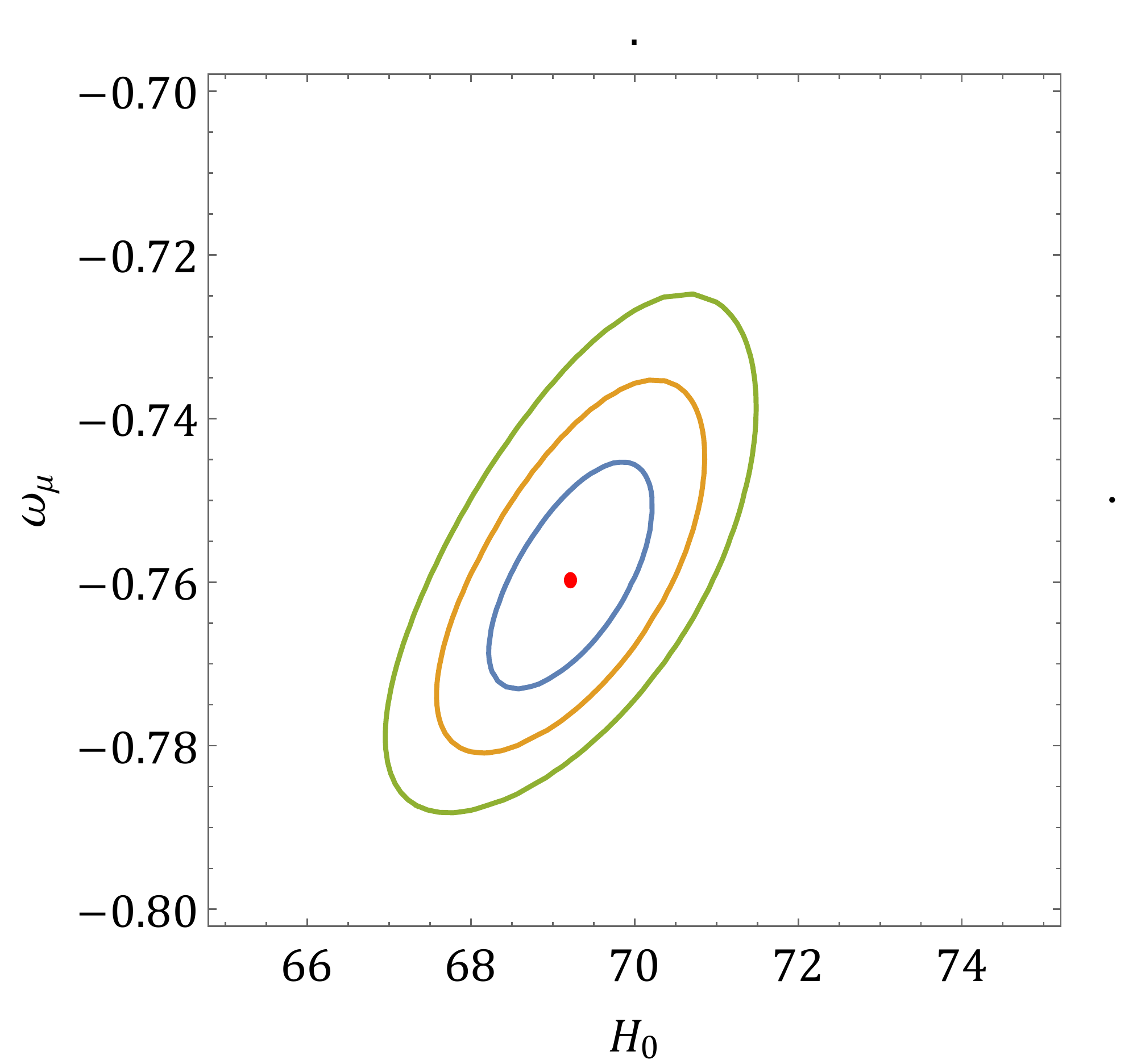}
	(c) \includegraphics[width=9cm,height=8cm,angle=0]{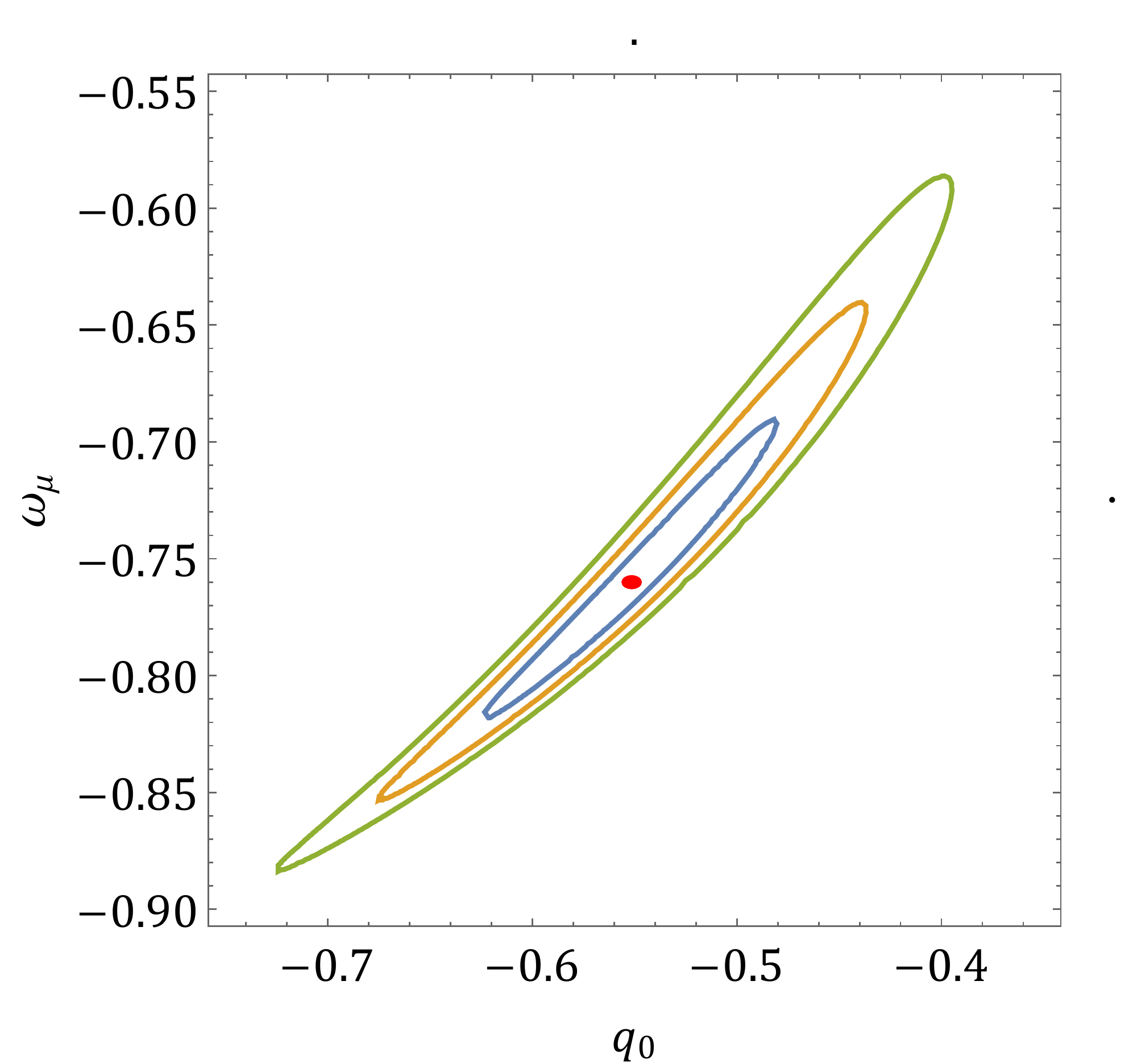}
	
	\caption{Figures (a), (b) and (c) are the 1$\sigma$, 2$\sigma$ and 3$\sigma$ confidence region plots for the pair of parameters ($H_0$, $q_0$ ), ($H_0$, $\omega_{\mu 0}$) and ($q_0$, $\omega_{\mu 0}$). The estimated value points  are red spotted.}
\end{figure}
\section{Distance Modulus and Apparent Magnitude for the Model:}
The luminosity distance ($d_L$) plays a very important parameter for observing distances of luminous objects like standard kindles, population I and II stars. The luminosity distance of the SN Ia supernovae was found to be  more than expected, which has given rise to the concept of the accelerating universe \cite{ref11,ref12}.   $d_L$ is given by:
\begin{equation}\label{21}
	\mu(z)= m_b - M = 5 Log d_L(z)+\mu_{0},
\end{equation}
where, $M$ and $ m_b$ are the absolute and apparent magnitudes, respectively. $\mu$ is called distance modulus. 
$ \mu_0 $ and $d_L$ are defined by 
\begin{equation}\label{22}
	\mu_0= 25+5 Log \big(\frac{c}{H_0} \big),
\end{equation}
and
\begin{equation}\label{23}
	d_L(z)=(1+z)H_0 \int_0^z{\frac{1}{H(z*)} dz*}.
\end{equation}
As we have already obtained expressions for H,  we can  determine  $d_L$ and  $\mu(z).$ We can also obtain expressions for apparent magnitude $m_b$ from these expressions. For this we use two important observations (i.) The absolute magnitude of all standard  candles are assumed to be the same, and (ii) The luminosity distance of very low red shift SNIa are approximated as $d_L=\frac{c z}{H_0}.$ Using these facts and taking a low red shift supernova with red shift $z=0.014$ and $m_b = 14.57$, we find $ M = -19.30.$ 
With this, we obtain the expression for the apparent magnitude as follows:
\begin{equation}\label{24}
	m_b = 5 Log(1+z) H_0 \int_0^z{\frac{1}{H(z*)} dz*}+5 Log \left(\frac{c}{H_0} \right)\mu_{0} + 5.70
\end{equation}

\section{Estimation of Model parameters from distance modulus and apparent data sets:} 
We use two standard data sets (i) 580 SN Ia data set of distance modulus (DM) in the red shift range  $ 0\leq z \leq 1.5 $ (ii) 66 pantheon data set of SN Ia apparent magnitude (AP) comprising 40 bin plus 16 high red shift data in the range  $0.014 \leq z \leq 2.26$. We consider theoretical results of distance modulus and apparent magnitude as obtained from Eqs. (\ref{21}) and (\ref{24}), as function of model parameters  $H_0$, $q_0$ and $\omega_{\mu 0}$. In these expressions we take red shifts from the data sets. Thus we form a parallel data set of theoretical results. With these, we form two chi squire functions separately for the DM and AP as follows:
\begin{equation}\label{25}
	\chi^{2}(H_{0}, q_{0},\omega_{\mu 0}) = \sum\limits_{i=1}^{580}\frac{[\mu th (z_{i},H_{0}, q_{0},\omega_{\mu 0}) - \mu ob(z_{i})]^{2}}{\sigma {(z_{i})}^{2}},
\end{equation}
and
\begin{equation}\label{26}
	\chi^{2}(H_{0}, q_{0},\omega_{\mu 0}) = \sum\limits_{i=1}^{66}\frac{[m_b th (z_{i}, H_{0}, q_{0},\omega_{\mu 0}) - m_{b} ob(z_{i})]^{2}}{\sigma {(z_{i})}^{2}},
\end{equation}

If the parameters $H_{0}, q_{0}$ and $\omega_{\mu 0}$ be given  in the range ( $ H_0$, 65$-$75),  ($q_0$, -0.60 $-$ -0.40) and ($\omega_{\mu 0}$, -0.90 $-$ -0.70), we  find  for the $\mu$ data set, $ ~H_0 = 70.0749,~q_0 =-0.628623~~ \text{and} ~~\omega _{\text{$\mu $0}}= -0.804005.$ 
$\text{for  minimum}~~ \chi^2 = 562.245 $, whereas for AP  data set it is
$H_0=71.7266,,~q_0 = -0.602399 ~\text{and} ~~\omega _{\text{$\mu $0}}= -0.798498$
$\text{for minimum}~~ \chi^2 = 71.861214.$ More over if we combine Hubble 77 data set and 580 distant modulus data set make a pool of 657 data set we get the following result:
 $ ~H_0 =  69.949,~q_0 = -0.61708~~ \text{and} ~~\omega _{\text{$\mu $0}}= -0.806833.$ 
$\text{for  minimum}~~ \chi^2 =615.311.$ We present the following table to display all of our statistical findings:
\begin{table}[H]
	\caption{ Statistical estimations for model parameters $H_0$, $q_0$ and $\omega _{\mu 0}$ }
	\begin{center}
		\begin{tabular}{|c|c|c|c|c|c|}
			\hline
			\\
			Datasets &~	 $H_{0}$~ &~~	 $q_{0}$~~&~~	 $\omega _{\text{$\mu $0}}$~&~~  $\chi^2$  ~ \\
			\\	
			\hline
			\\
		77	$OHD$~H  &~ $ 69.2163 $ ~ &~ $-0.551698~ $~~&~~-0.759826~~&~~52.0805\\ 
			\\	
			\hline
			\\
		580	D.M	$\mu$ &~  $ 70.0749$	 ~&~ $-0.628623~ $~~&~~ -0.804005~~&~~562.245 \\
			\\	
			\hline
			\\
		66	A.M	$m_b$ &~  $71.7266$	 ~ &~ $-0.602399~ $ ~~&~~ -0.798498~~&~~ 71.861214\\
			\\	
			\hline
			\\
		657	$OHD$ H  +	D.M	$\mu$ &~  $ 69.949$	 ~ &~  $  -0.61708~$ ~~&~~ -0.806833~~&~~615.311\\
			\\	
			\hline
				
		\end{tabular}
	\end{center}
\end{table}
 Figure $3(a)$ describes the growth of  distant modulus $\mu$ over red shift $ `z'$. The distant modulus  is increasing function of red shift which means that in the past distant modulus was more. It is gradually decreasing over time. The figure also shows that theoretical graph passes near by through the 580 vertical error lines whose middle points  are observed values at different red shifts. Figure $3(b)$, $3(c)$ and $3(d)$ are the 1$\sigma$, 2$\sigma$ and 3$\sigma$ confidence region plots for the pair of parameters ($H_0$, $q_0$ ), ($H_0$, $\omega_{\mu 0}$) and ($q_0$, $\omega_{\mu 0}$). The estimated value points $(H_0= 70.0749, q_0 = -0.628623)$, $(H_0= 70.0749, \omega _{\text{$\mu $0}} = -0.804005)$ and $(q_0 =-0.628623, \omega _{\text{$\mu $0}} =-0.804005)$ are red spotted. These are just there to show that our estimated values are within the statistically specified regions.
\begin{figure}[H]
	(a)	\includegraphics[width=8cm,height=7cm,angle=0]{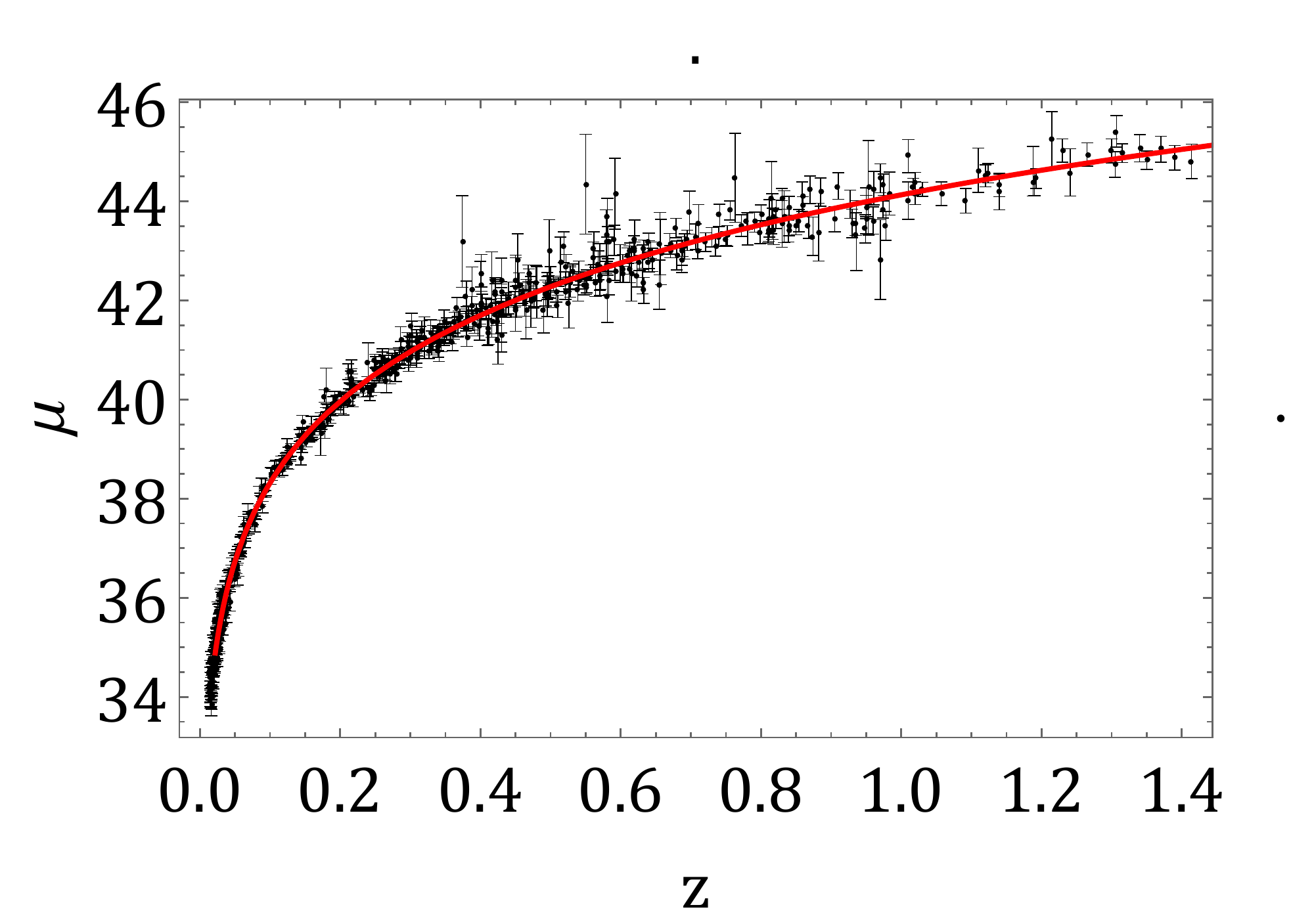}
	(b) \includegraphics[width=8cm,height=7cm,angle=0]{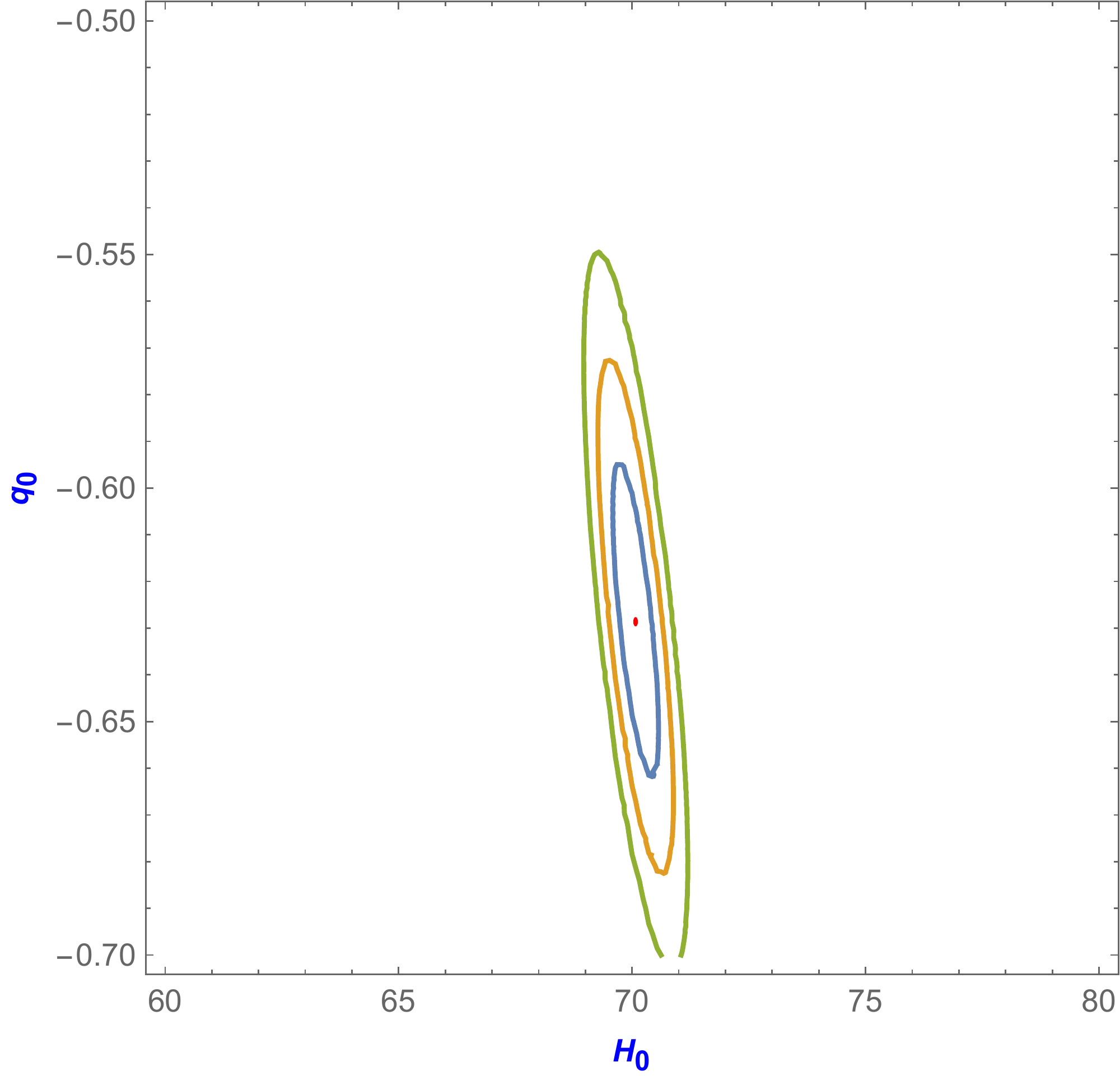}
	(c) \includegraphics[width=8cm,height=7cm,angle=0]{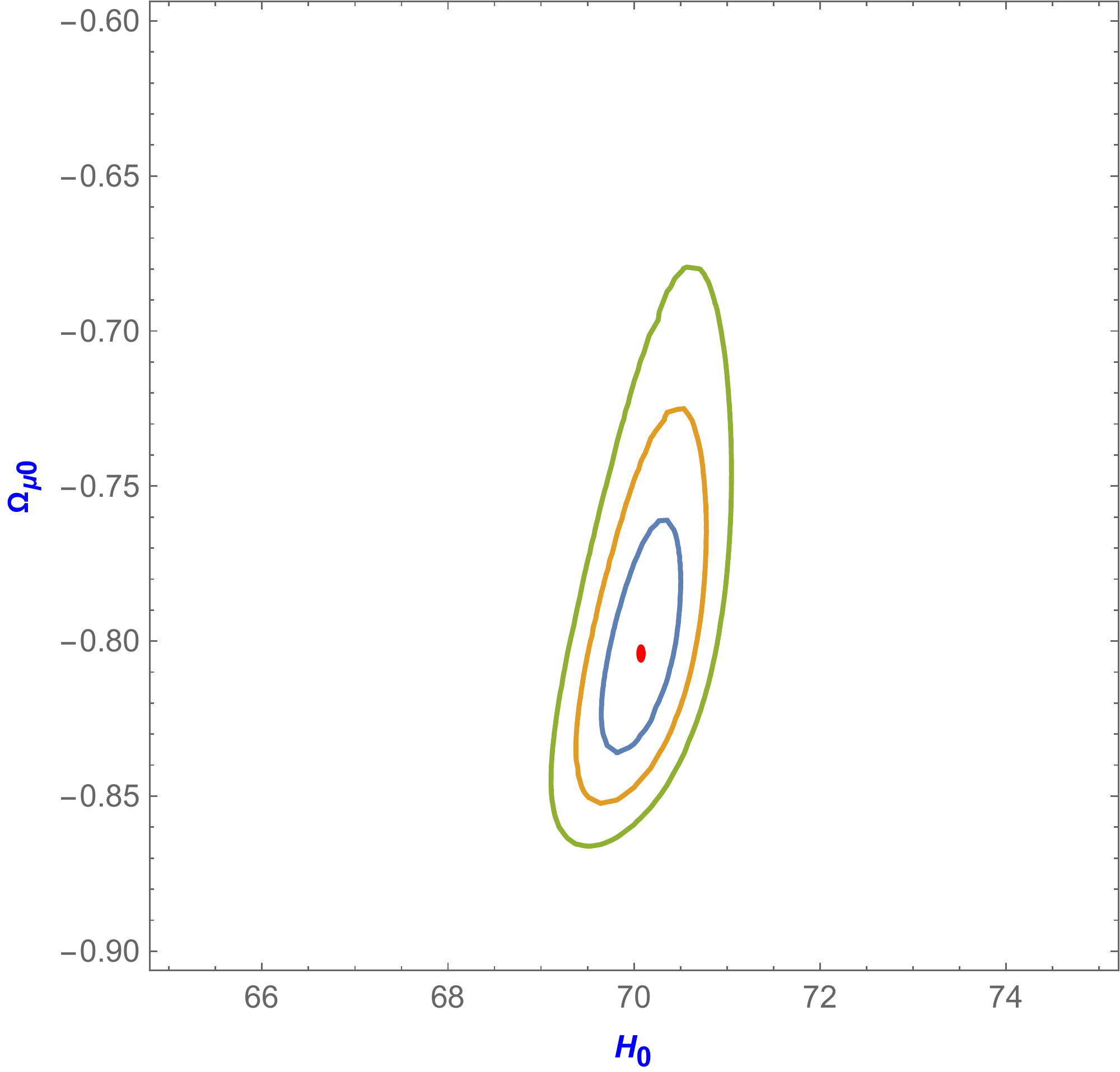}
	(d)	\includegraphics[width=8cm,height=7cm,angle=0]{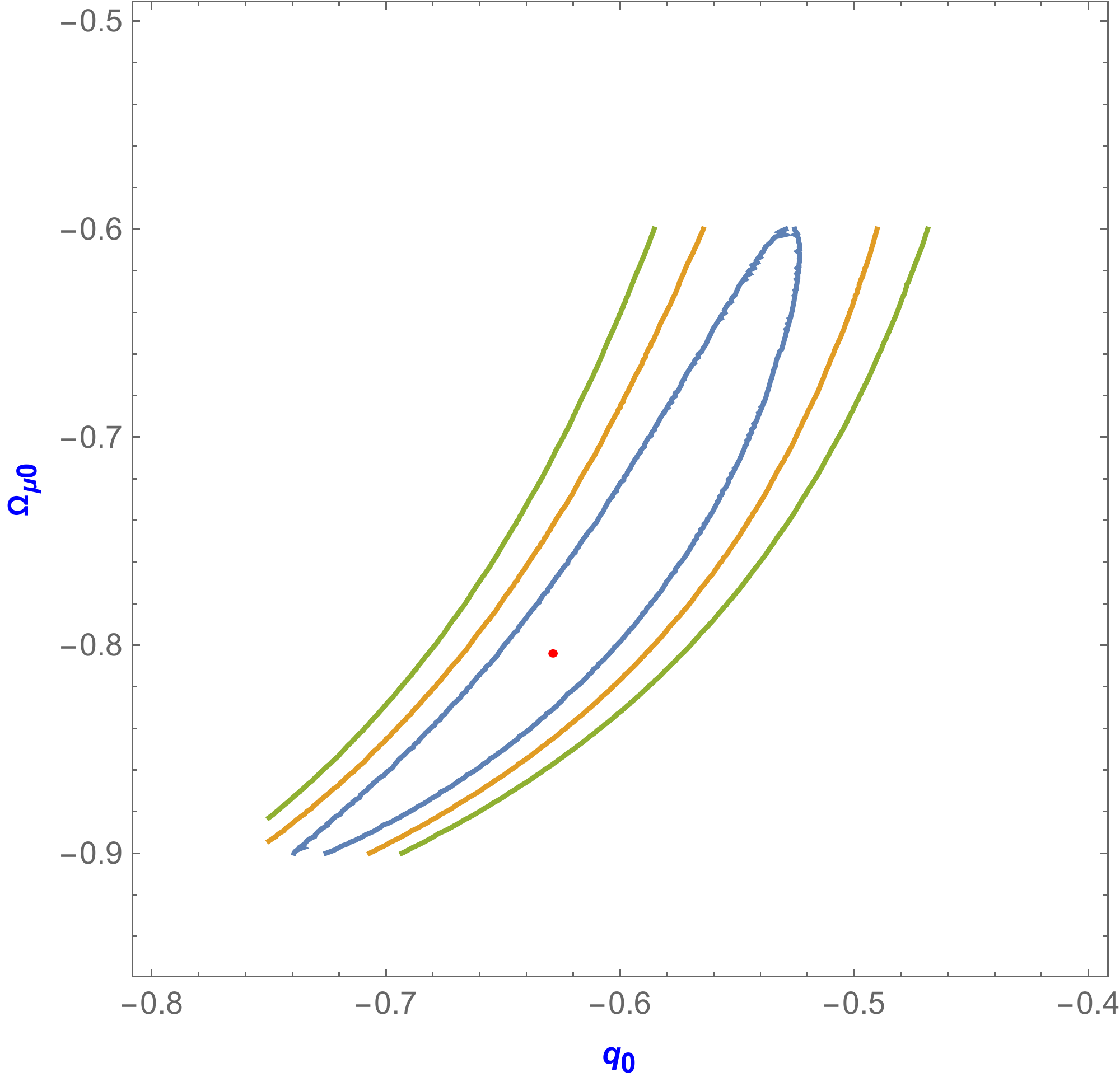}\\
\caption{Figure (a)  is the error bar plot  for   $\mu$ over  $ z$. Figures (b), (c) and (d) are the 1$\sigma$, 2$\sigma$ and 3$\sigma$ confidence region plots for  ($H_0$, $q_0$ ), ($H_0$, $\omega_{\mu 0}$) and ($q_0$, $\omega_{\mu 0}$). The estimated value points $(H_0= 70.0749,q_0 = -0.628623)$, $(H_0= 70.0749, \omega _{\text{$\mu $0}} = -0.804005)$ and $(q_0 =-0.628623,\omega _{\text{$\mu $0}} =-0.804005)$ are red spotted. }
\end{figure}
 Figures $4(a)$, $4(b)$ and $4(c)$ are likelihood probability curves for Hubble $H_0$, deceleration 
$ q_0$ and equation of state $ \omega_{\mu 0}$ parameters. Estimated values  on the basis of 580 SNIa D.M.($\mu$) data set $H_0= 70.0749,  q_0 = -0.628623, \omega _{\text{$\mu $0}} = -0.804005$  are at the peak.
\begin{figure}[H]	
	(a) \includegraphics[width=8cm,height=7cm,angle=0]{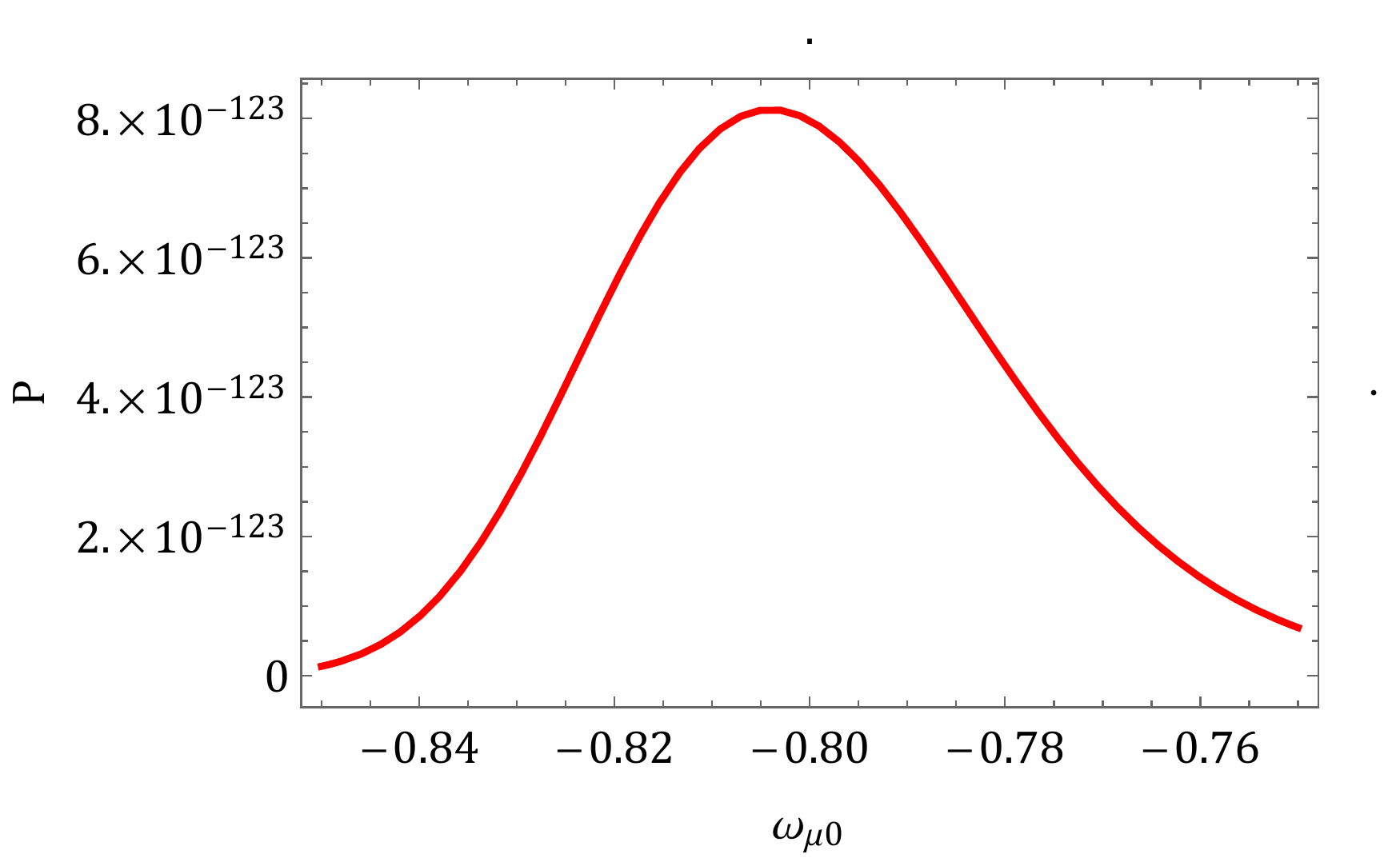}
	(b)\includegraphics[width=8cm,height=7cm,angle=0]{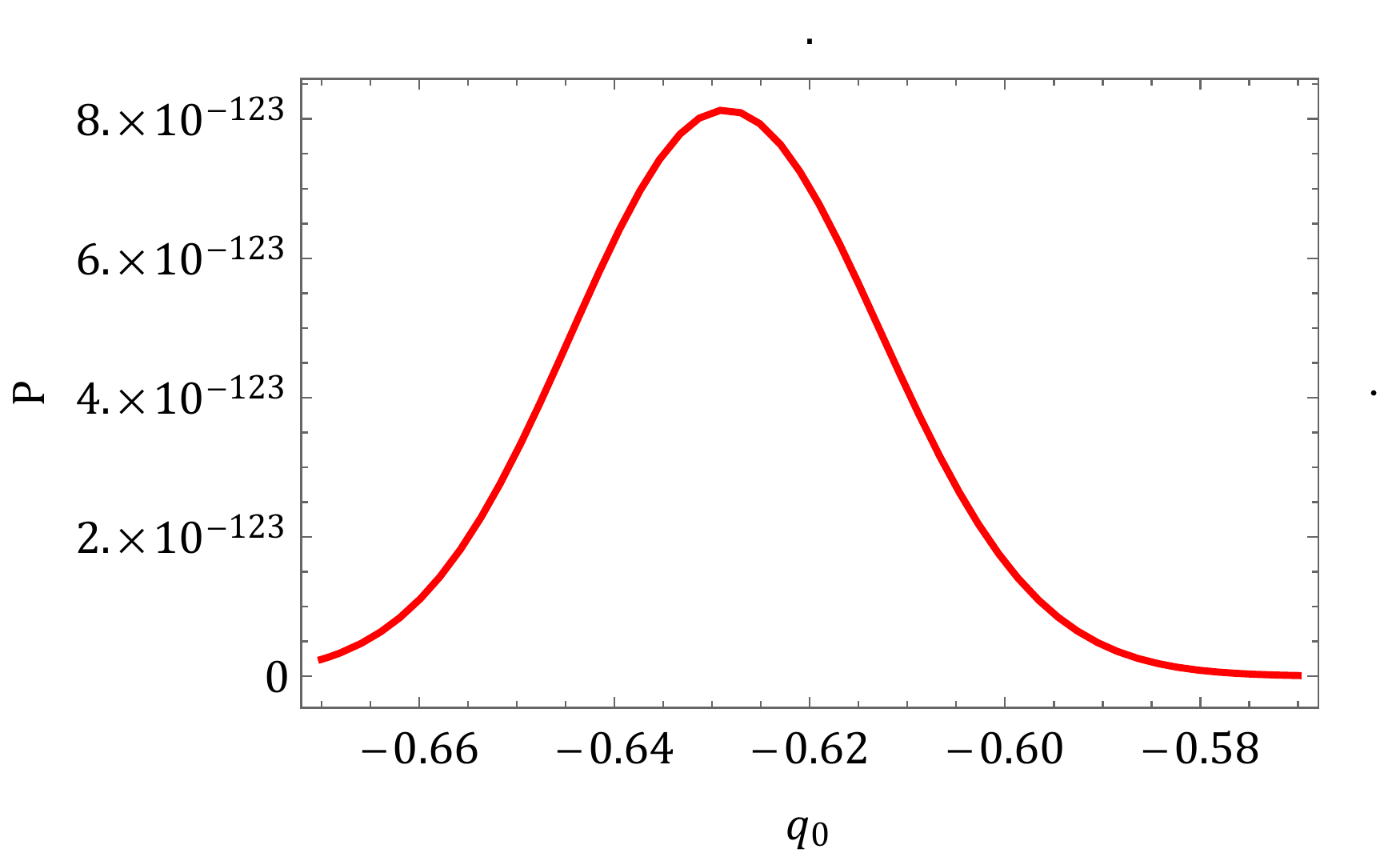}
	(c)\includegraphics[width=8cm,height=7cm,angle=0]{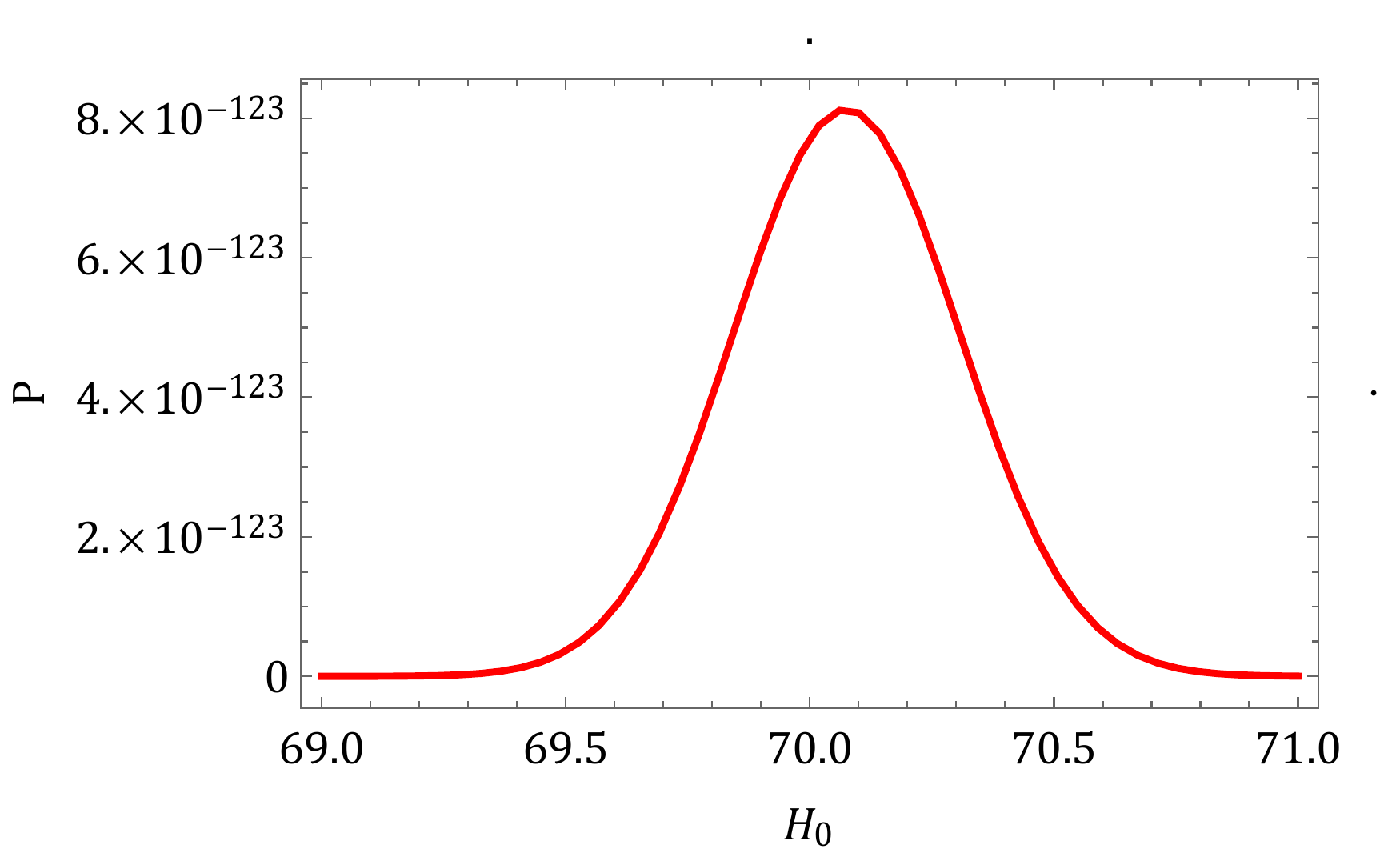}
	\caption{Figures (a), (b) and (c)  are likelihood probability curves for   $H_0$,  $ q_0$ and $ \omega_{\mu 0}$ parameters. Estimated values on the basis of 580 SNIa DM ($\mu$) data set $H_0= 70.0749, q_0 = -0.628623, \omega _{\text{$\mu $0}} = -0.804005$  }
	\end{figure}
Figure $5(a)$ describes the growth of AM $m_b$ over red shift $ `z'$. The  AM  is increasing function of red shift which means that in the past DM was more. It is gradually decreasing over time. The figure also shows that theoretical graph passes near by through the 66 vertical error lines whose middle points  are observed values of AP at different red shifts. Figure $5(b), 5(c) ~\text{and}~ 5(d)$ are the 1$\sigma$, 2$\sigma$ and 3$\sigma$ confidence region plots for the pair of parameters ($H_0$, $q_0$ ), ($H_0$, $\omega_{\mu 0}$) and ($q_0$, $\omega_{\mu 0}$). The estimated value points $(H_0= 71.7266, q_0 = -0.602399)$, $(H_0= 71.7266, \omega _{\text{$\mu $0}} = -0.798498)$ and $(q_0 =-0.602399,\omega _{\text{$\mu $0}} =-0.798498)$ are red spotted.
\begin{figure}[H]	
	(a) \includegraphics[width=8cm,height=7cm,angle=0]{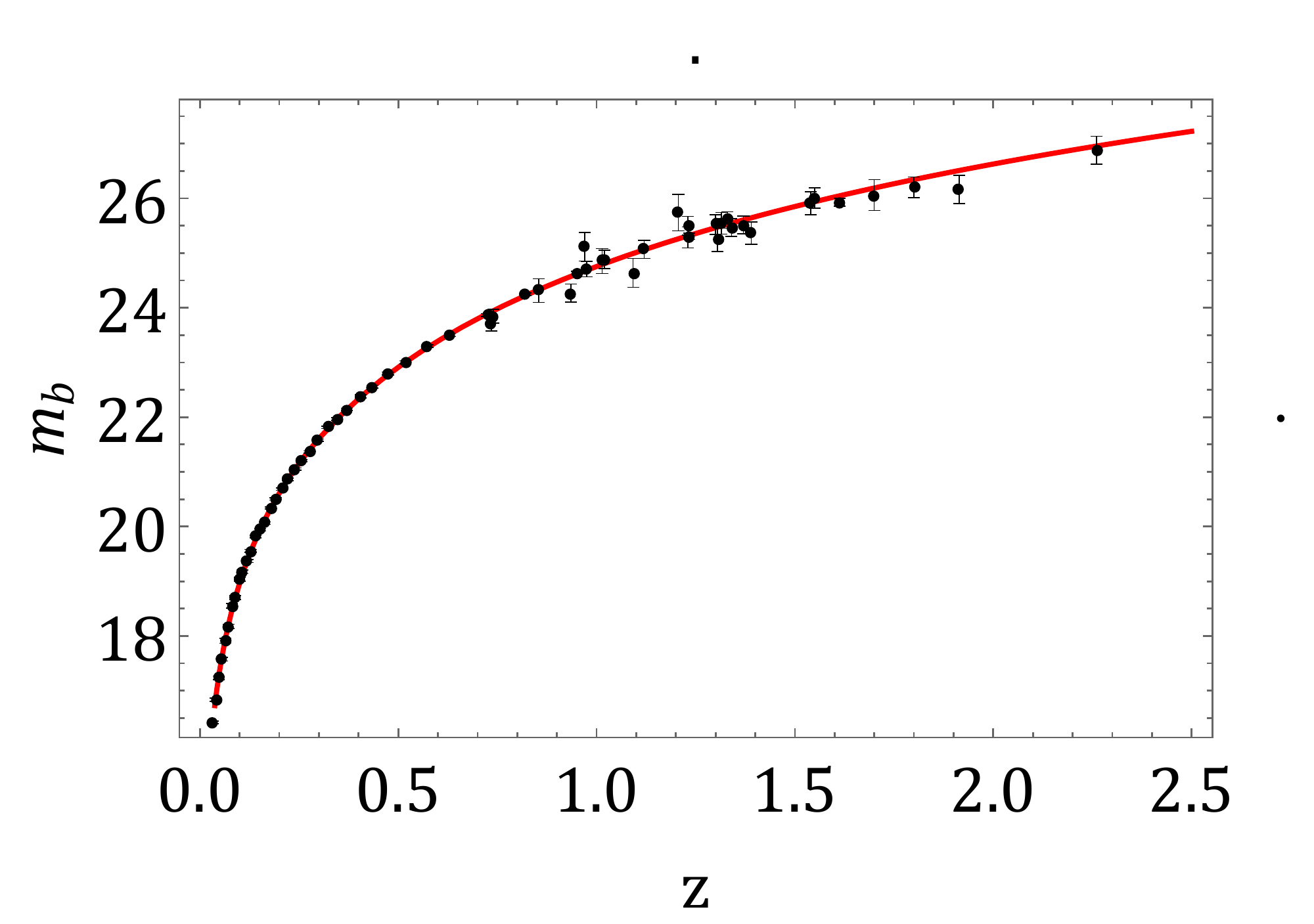}
	(b)\includegraphics[width=8cm,height=7cm,angle=0]{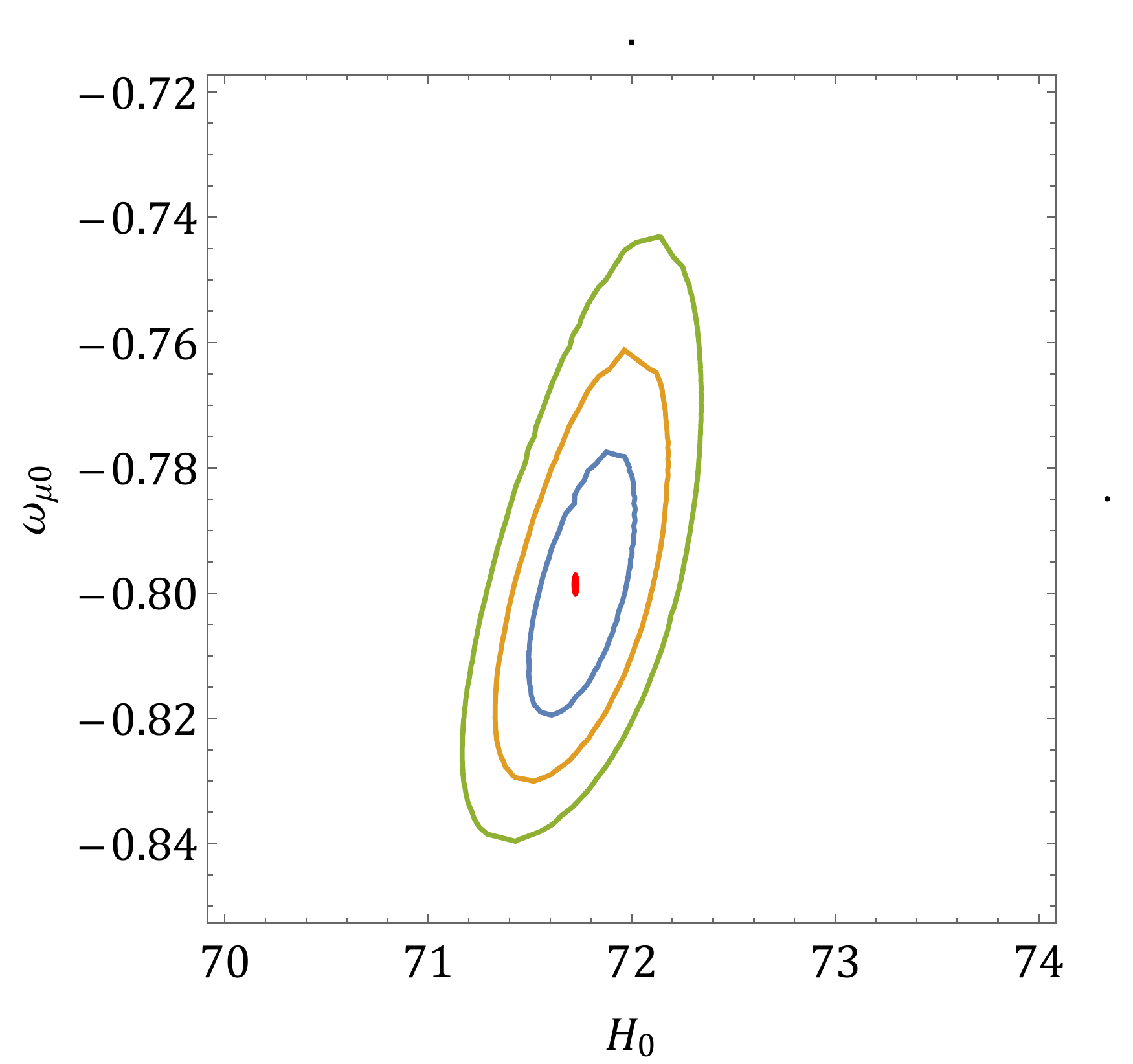}
	(c)\includegraphics[width=8cm,height=7cm,angle=0]{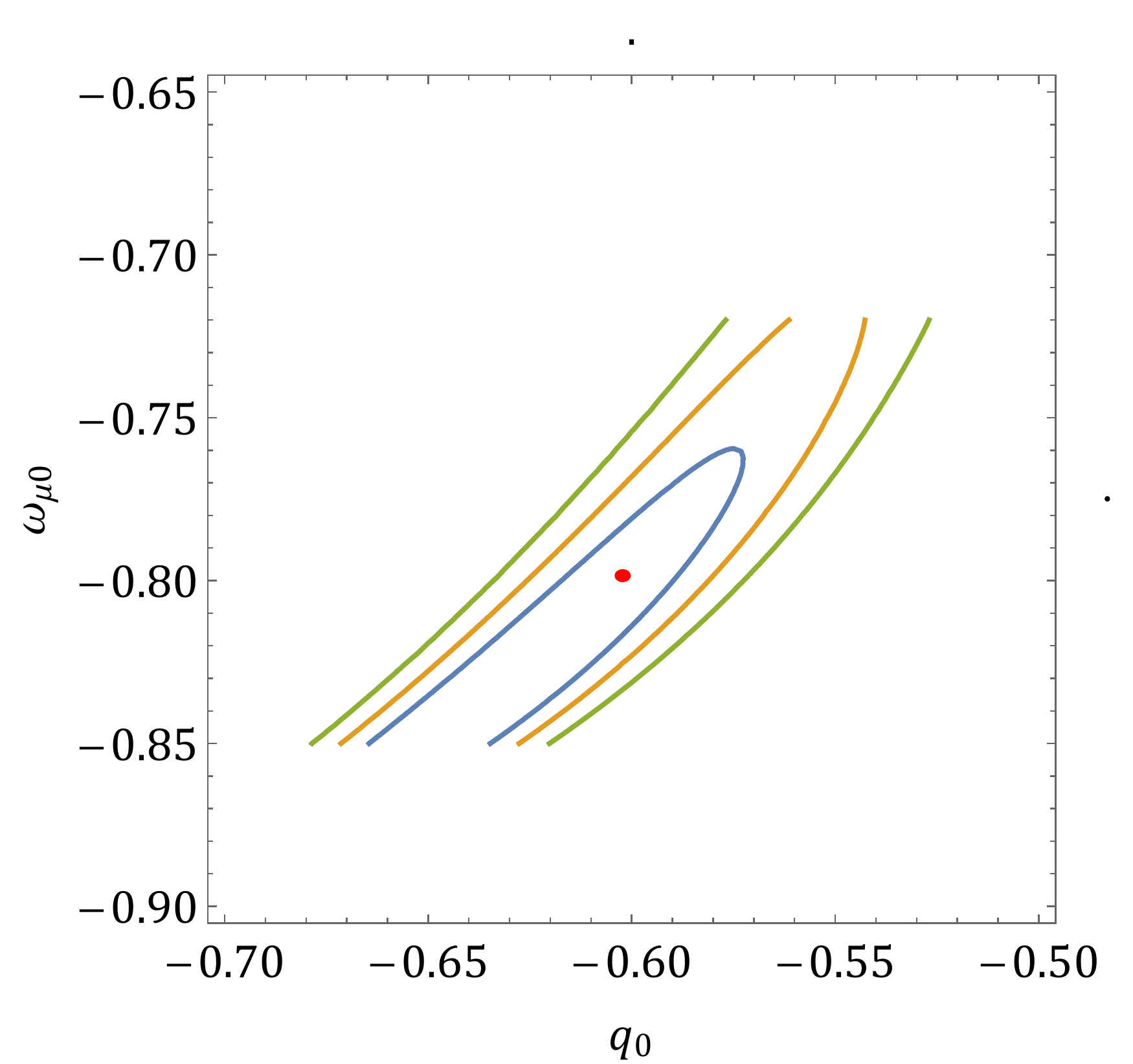}
	(d) \includegraphics[width=8cm,height=7cm,angle=0]{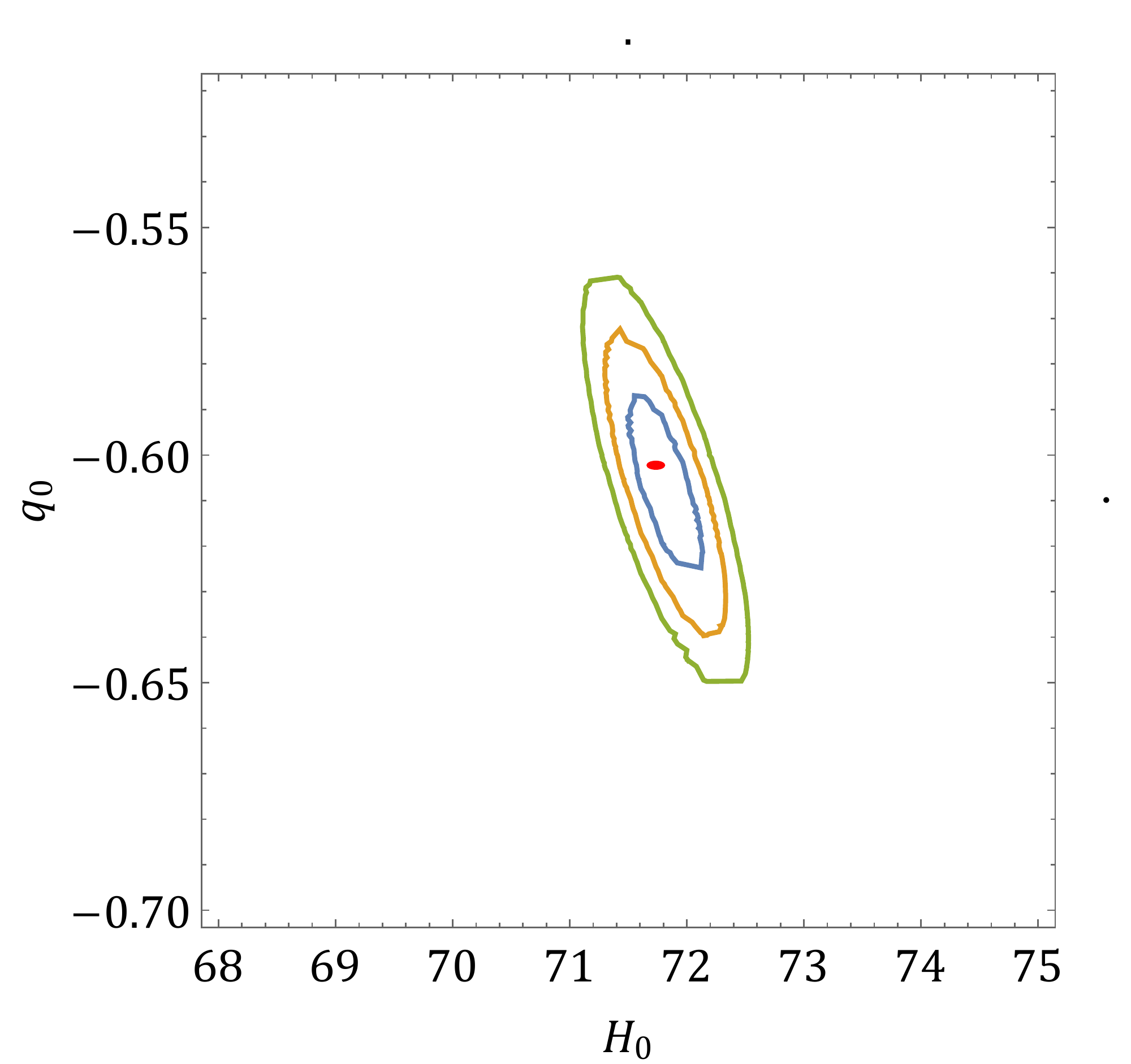}
	\caption{Figure (a)  is the error bar plot  for  AM over  $ z$. Figures (b), (c) and (d) are the 1$\sigma$, 2$\sigma$ and 3$\sigma$ confidence region plots for  ($H_0$, $q_0$ ), ($H_0$, $\omega_{\mu 0}$) and ($q_0$, $\omega_{\mu 0}$). The estimated value points$(H_0= 71.7266, q_0 = -0.602399)$, $(H_0= 71.7266, \omega_{\text{$\mu $0}} = -0.798498)$ and $(q_0 =-0.602399,\omega _{\text{$\mu $0}} =-0.798498)$ are red spotted.}
\end{figure}
Figures $6(a)$, $6(b)$ and $6(c)$ are likelihood probability curves for Hubble  $H_0$, deceleration $ q_0$ and equation of state $ \omega_{\mu 0}$ parameters. Estimated values on the basis of 66 Pantheon apparent modulus data set $H_0= 71.7266, q_0 =-0.602399, \omega_{\text{$\mu $0}} = -0.798498$  are at the peak.

\begin{figure}[H]	
	
	(a)\includegraphics[width=8cm,height=7cm,angle=0]{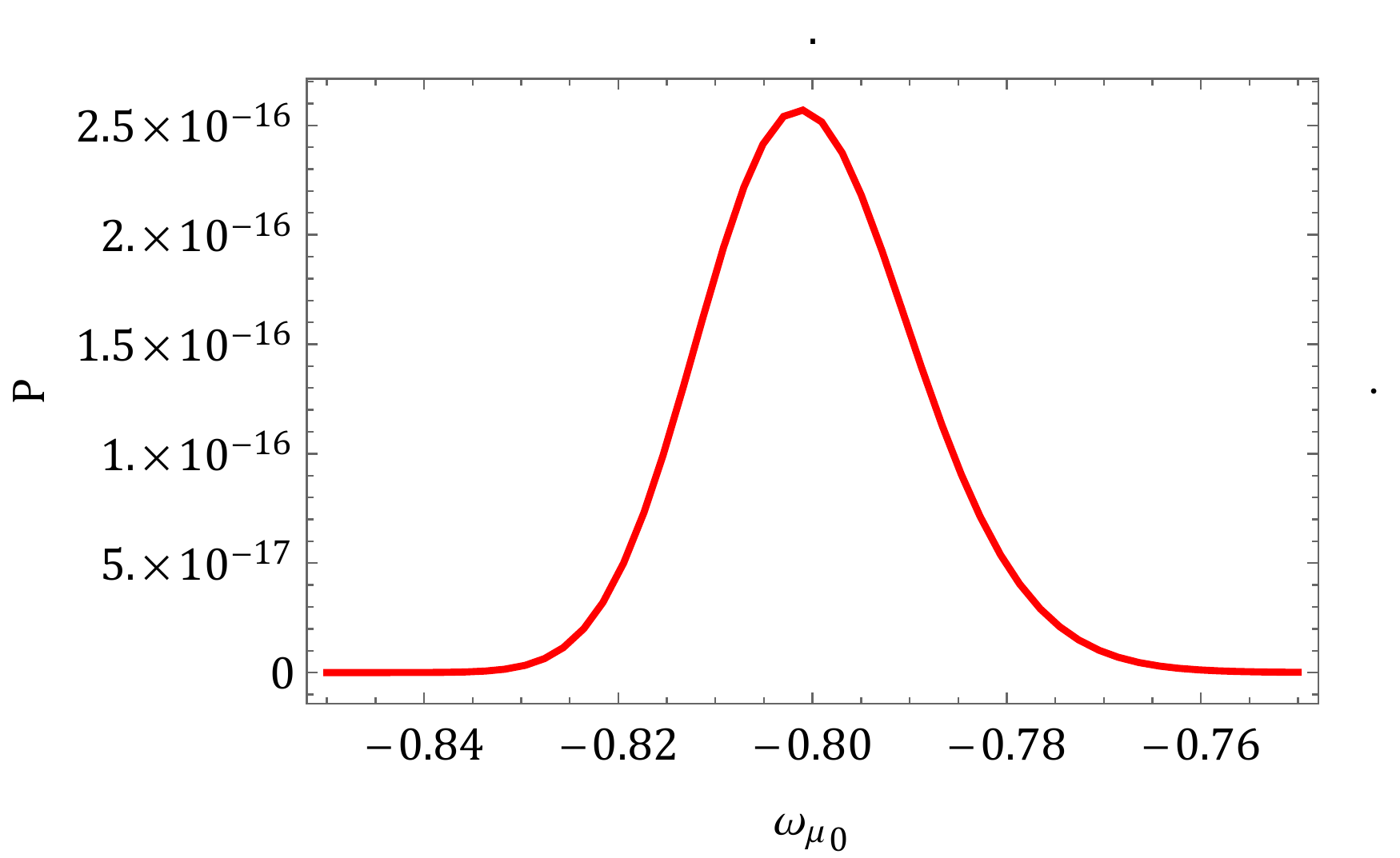}
	(b)\includegraphics[width=8cm,height=7cm,angle=0]{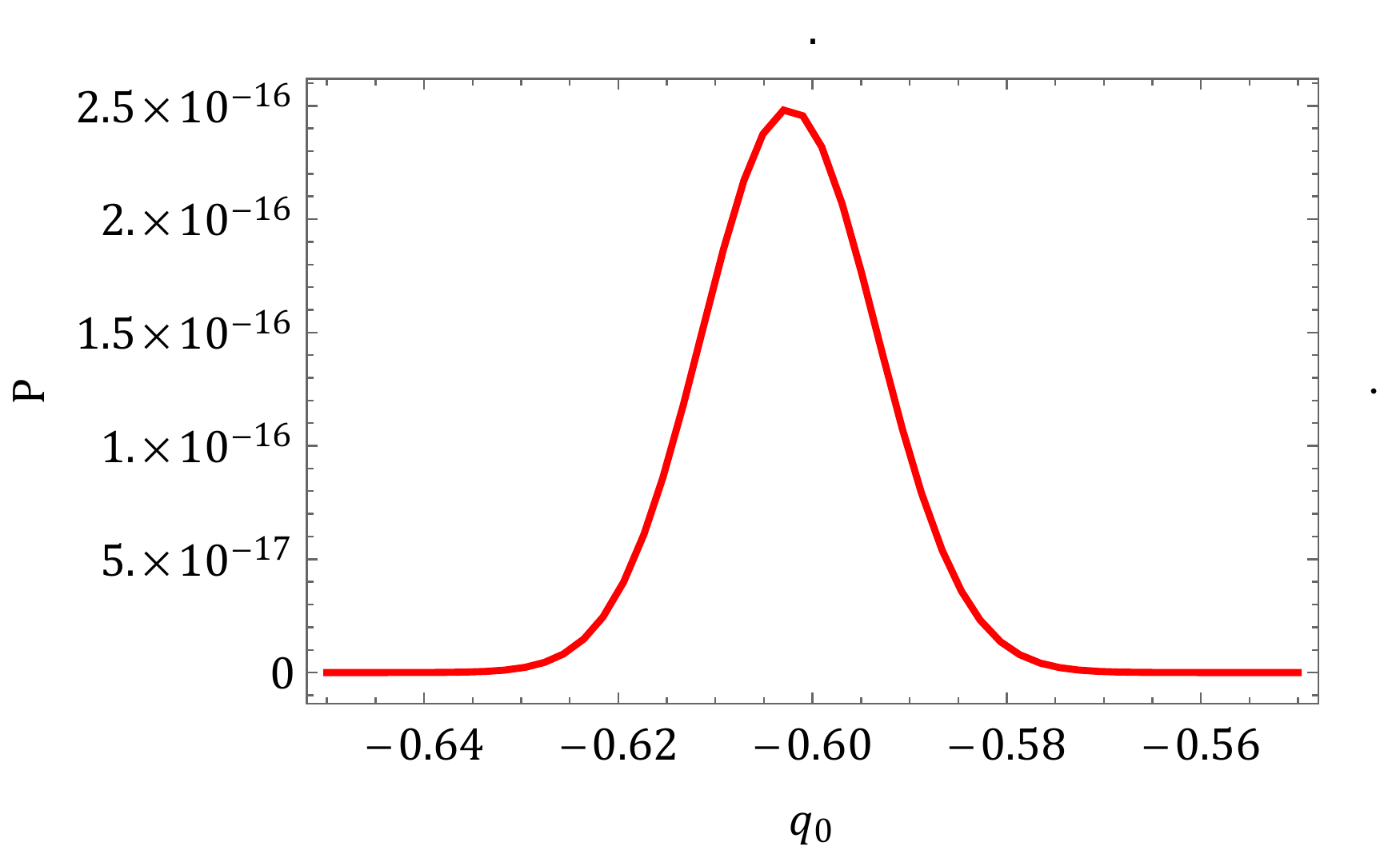}
	(c) \includegraphics[width=8cm,height=7cm,angle=0]{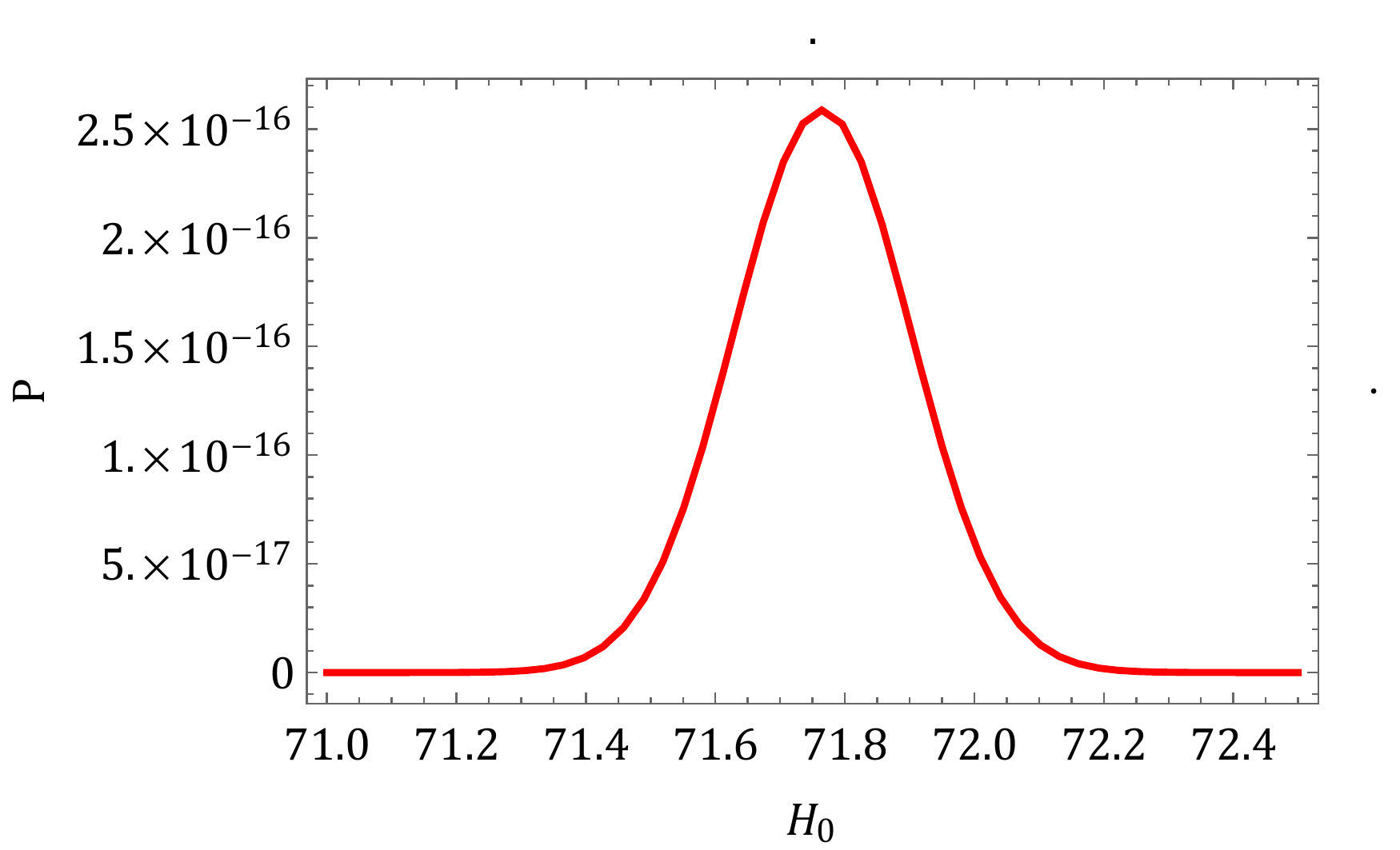}
	\caption{Figures (a), (b) and (c)  are likelihood probability curves for   $H_0$, $ q_0$ and  $ \omega_{\mu 0}$ parameters. Estimated values  $H_0= 71.7266,q_0 =-0.602399, \omega _{\text{$\mu $0}} = -0.798498$, which are obtained on the basis of 66 Pantheon apparent modulus data set,  are at the peaks.}
\end{figure}

\subsection{Determination of the present values  of Energy parameters 	$\Omega_{m 0}$ and 
	$\Omega _{\text{$\mu $0}}$:}
In sections $4$ and $6$, we have estimated the present values of  $H_0$, $ q_0$, and $ \omega_{\mu 0}$  on the basis of four data sets which are described in Table-1. From Eqs. (\ref{13}) and (\ref{14}), we obtain values of  $\mu=8\pi \lambda$ and  $\omega_m$ for matter. From Eq.
 (\ref{9}), we get energy parameters $\Omega_{\mu 0}$ and
$\Omega_{m  0}$ as follows: 
 
\begin{equation}\label{27}
\Omega_{m 0} = \frac{8\pi \rho}{3 H_0^2} = \frac{1}{1+\mu(3-\omega_{m 0})};~~	\Omega _{\text{$\mu $0}} = 1 - \Omega_{m 0}
\end{equation}	
\begin{table}[H]
	\caption{Present Value of Energy Parameters $\Omega_m$ and $\Omega_{\mu_0}$}
	\begin{center}
		\begin{tabular}{|c|c|c|c|c|c|}
			\hline
			\\
			Datasets &~	 $\mu$~ &~~	 $\omega_m$~~&~~$\Omega_{m 0}$	~&~~ $\Omega _{\text{$\mu $0}}$  ~ \\
			\\	
			\hline
			\\
			77	$OHD$~H  &~ $0.620112 $ ~ &~ $-0.571151~ $~~&~~0.311089~~&~~0.688911\\ 
			\\	
			\hline
			\\
			580	D.M	$\mu$ &~  $0.582203$ ~&~ $-0.642996~ $~~&~~0.320414~~&~~0.679586 \\
			\\	
			\hline
			\\
			66	A.M	$m_b$ &~  $0.437467$	 ~ &~ $-0.633883~ $ ~~&~~0.386145~~&~~0.613855\\
			\\	
			\hline
			\\
			657	$OHD$ H  +	D.M	$\mu$ &~  $ 0.428288$~ &~ $-0.647688~$ ~~&~~0.39028~~&~~0.60972\\
			\\	
			\hline
			
		\end{tabular}
	\end{center}
\end{table}	
In this table, we find that at present $\Omega_{\mu}$ is dominant and it is nearly in the ratio 1:3 or 2:3 as per different observed  data sets. We note that $\Omega_\mu$  is  due to the curvature and energy momentum dominance in $f(R,T)$ gravity.

\section{Plots of Deceleration, Jerk and Snap Parameters:}

 The jerk (j) and snap (s)  parameters  are related to the third and fourth order derivatives of the scale factor. They play a very important role in examining the instability of a cosmological model. The jerk  also plays a role in statefinder diagnostics. They are defined as:  $j=\frac{\dddot{a}}{a H^3}$ and  $s= -\frac{\ddddot{a}}{a H^4}. $
j in the terms of q can be written as: 
\begin{equation}{\label{28}}
	j(z) = q(z) +2 {q(z)}^2 + (1+z) \frac{d q(z)}{d z}.
\end{equation}
whereas s in terms of q and j is computed as:
\begin{equation}{\label{29}}
	s(z) = ( 3 q(z) +2 ) j(z) + \frac{d j(z)}{d z} (1+z)
\end{equation}
Having estimated  the present values of  $H_0$, $q_0$, $\omega_{\mu 0}$ and $\mu$ in sections $4$ and $6$, we can plot and analyse q, j and s which are presented in the following figures.

\begin{figure}[H]	
	
	(a)\includegraphics[width=8cm,height=7cm,angle=0]{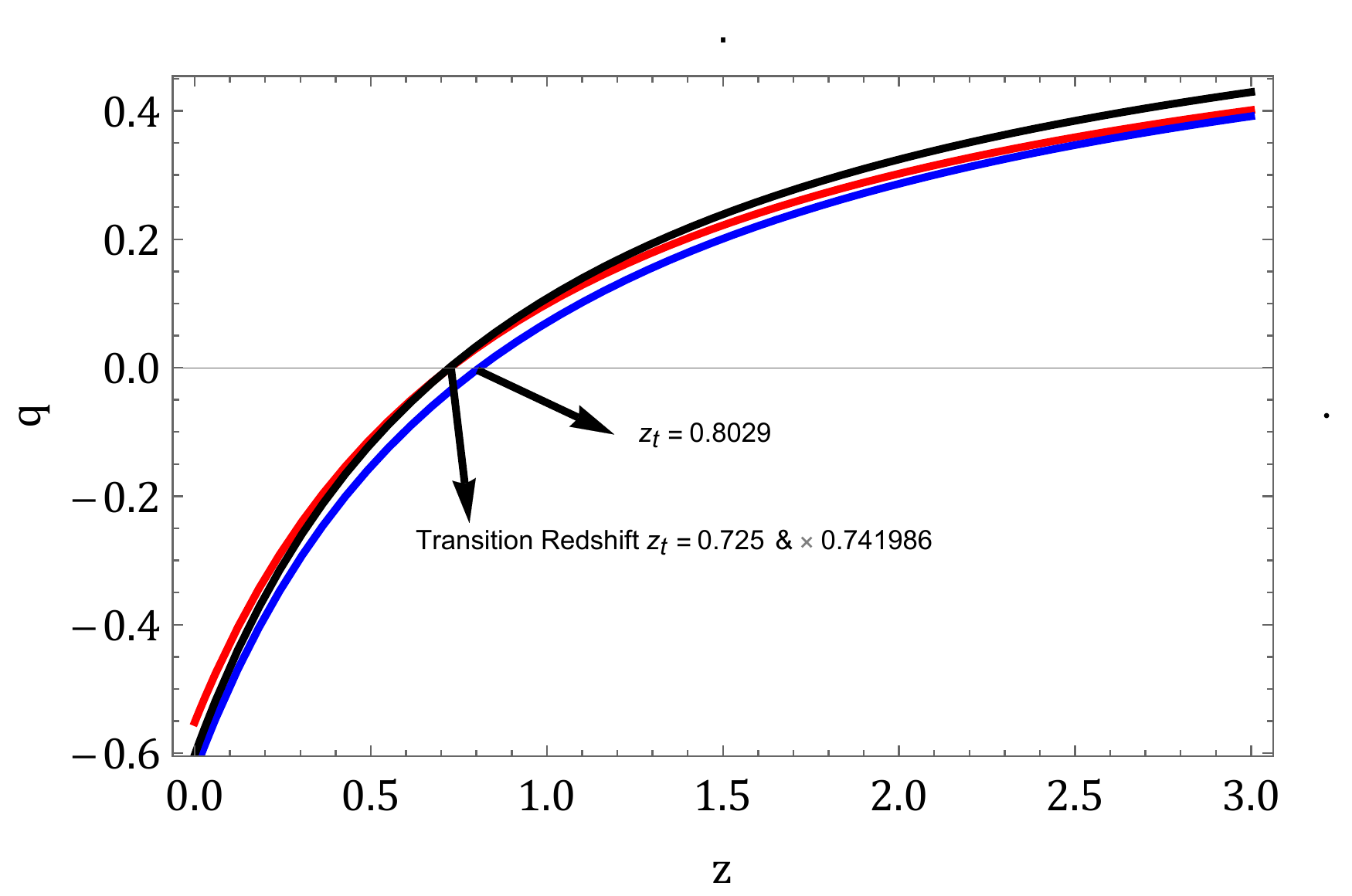}
	(b)\includegraphics[width=8cm,height=7cm,angle=0]{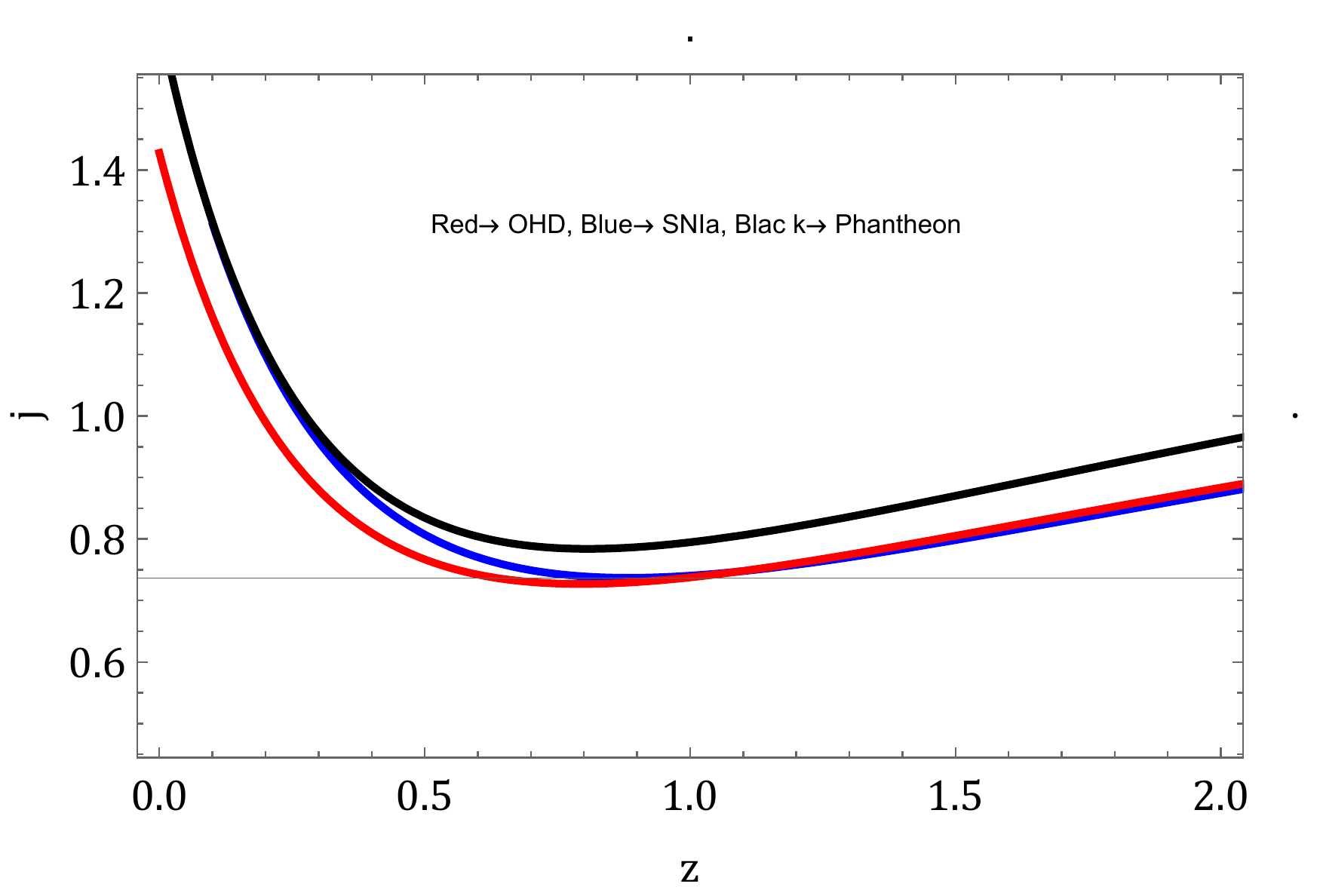}
	(c) \includegraphics[width=8cm,height=7cm,angle=0]{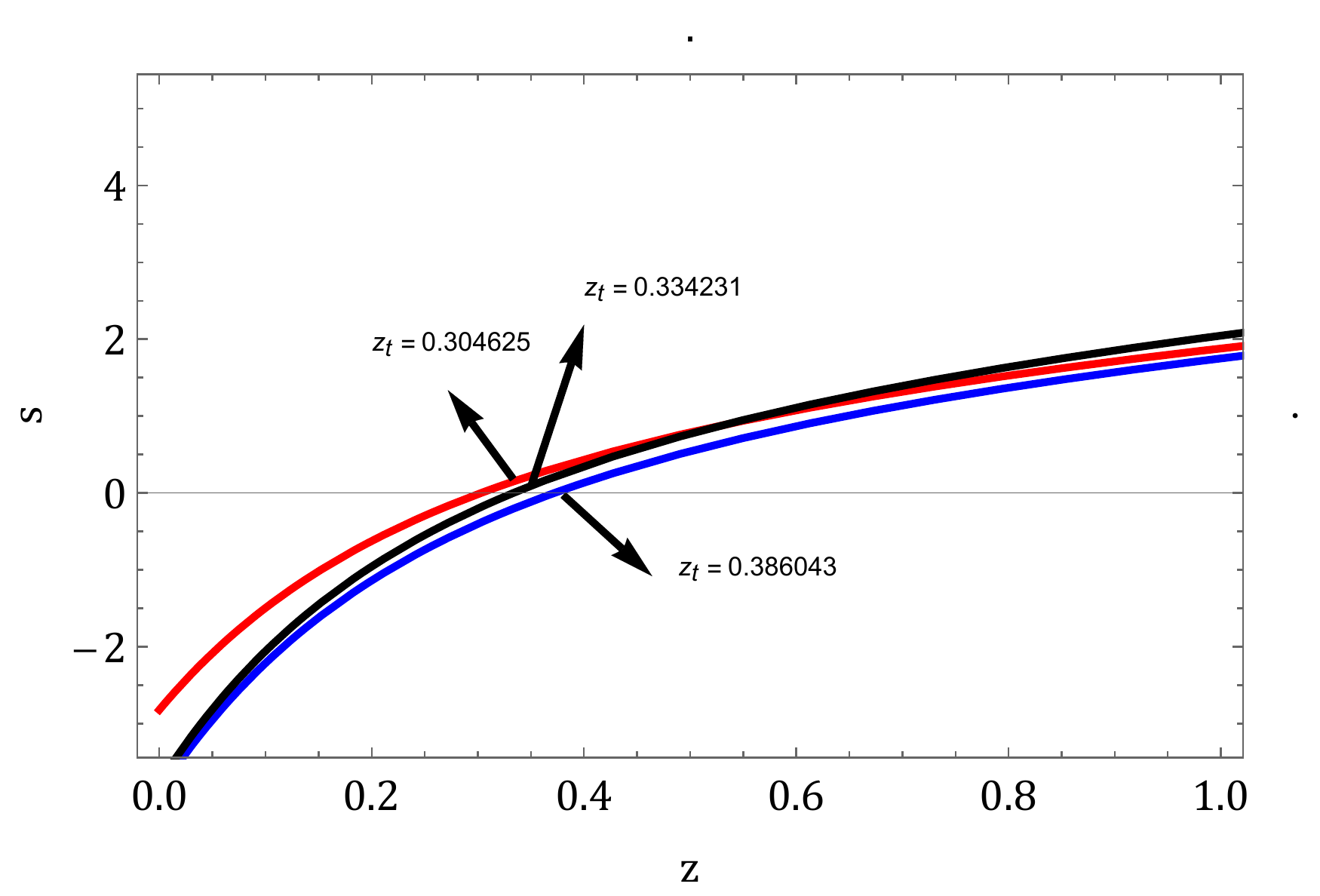}
	\caption{Figures (a), (b) and (c) are plots for  $q$, $j$ and $s$. Each plot contains three curves  which corresponds to the three data sets described in sections (4) and (6). Transition red shifts for $q$ and $s$ are displayed. $j$ is always positive.}
\end{figure}

We make following observations from  the  three plots of  Figure $7$
\begin{itemize}
	\item q  and s  show a transition from negative to positive, which indicates that the universe is accelerating at present, and it was decelerating in the past.  
	\item The Transition red shifts are well within the observational results 
	\item j is always positive. Its present values are 1.42783, 1.6416 and 1.64327
	as per the OHD, SNIa and pantheon data sets. These values are greater than one, which shows that this model behaves differently from the $\Lambda$CDM concordance model where j=1.
\end{itemize}

\section{Statefinder Diagnostic:}
In this important section, we use a very useful diagnostic technique given by Sahni {\it et al.} \cite{ref41}. They have used a pair of statefinder parameters $(r, s)$ depending on the scale factor $a$. The purpose is to differentiate cosmological models from the standard  $\Lambda$CDM concordance model. The techniques also describe the evolution of the models in the sense that it indicates through which phases the model has passed through.
$s(z)$ is defined as follows:

\begin{equation}\label{30}
	s(\text{z})=\frac{r(z)-1}{3 \left(q(z)-\frac{1}{2}\right)},
\end{equation}
where $r$ is the jerk parameter. \\

We plot two parametric curves, one with $s$ against $r$,  and the other with $q$ against $r$.
\begin{figure}[H]	
	(a)\includegraphics[width=8cm,height=7cm,angle=0]{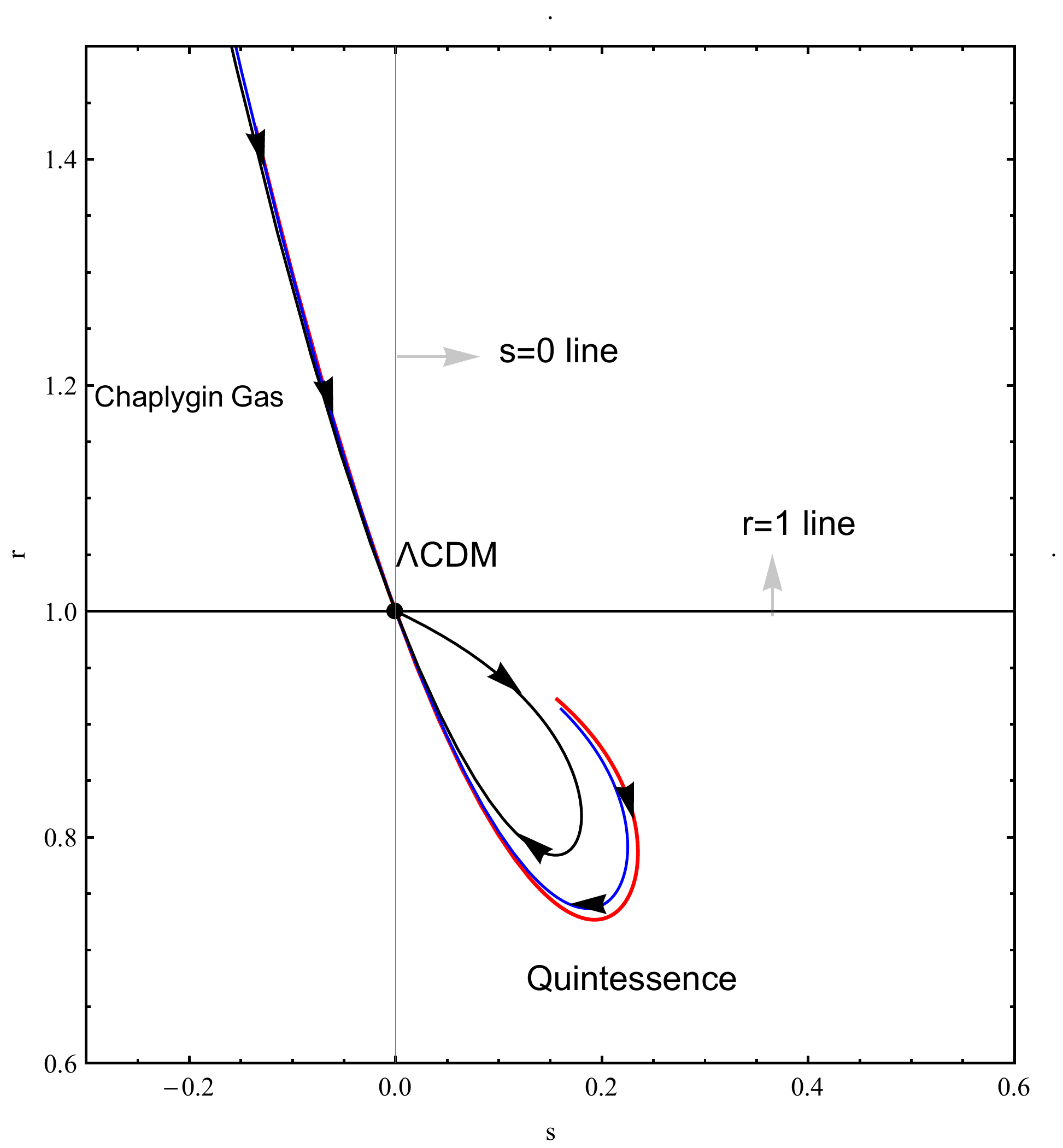}
	(b)\includegraphics[width=8cm,height=7cm,angle=0]{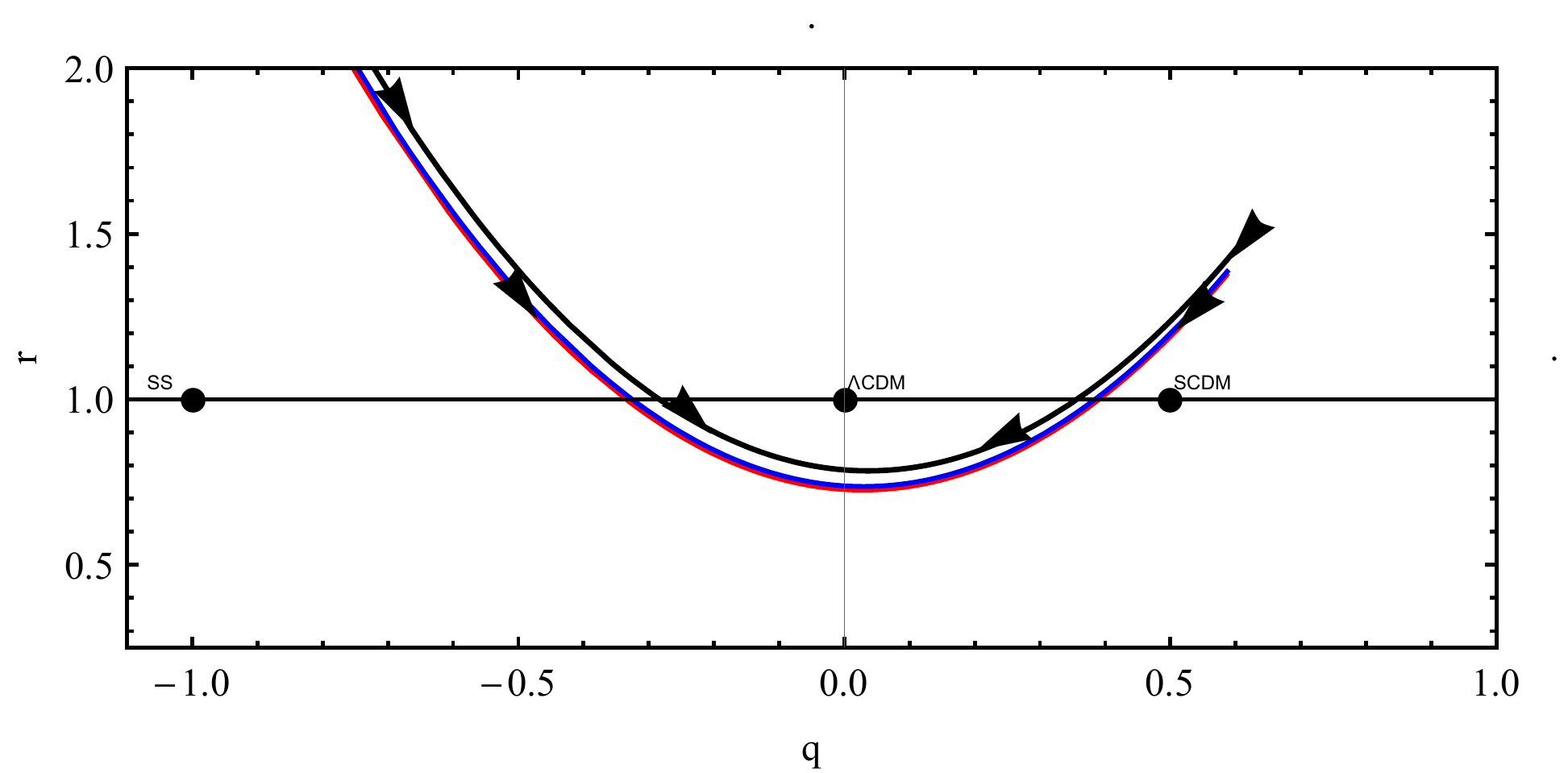}
	\caption{Figures (a) and (b) are  plots of $s$ versus $r$ and $q$ versus $r$. These show that at present our model is in quintessence. In the past too, it was not Einstein De sitter as reflected in the figures.}
	\end{figure}
	We make following observations as reflected in the   two plots of Figure $8$:
	\begin{itemize}
		\item Fig (a) shows that the three parametric (s,r) curves based on the the estimated values of the OHD, SNIa and phantom data sets meet at $\Lambda$CDM point(1,0) from both directions, i.e., from the Chaplygin gas model to quintessence. 
		\item  At present, our model is in quintessence. 
		\item The Fig. (b) also tells us that our model does not approximate the $\Lambda$CDM model, nor was it the Einstein-Desitter model in the past. 
	\end{itemize}

\section{Density and Pressure in the Model:}

From Eqs. (\ref{8}) and (\ref{9}), we get, $\rho$, $p$ and  $\omega_m = \frac{p}{\rho} $, for the matter as:

\begin{equation}\label{31}
8\pi	\rho =\frac{H^2 (2 \mu  (q+4)+3)}{(2 \mu +1) (4 \mu +1)}
\end{equation}

\begin{equation}\label{32}
	8\pi p = \frac{H^2 ((6 \mu +2) q-1)}{(2 \mu +1) (4 \mu +1)}
\end{equation}
and 
\begin{equation}\label{33}
	\omega_m =\frac{p}{\rho}=\frac{(6 \mu +2) q-1}{2 \mu  (q+4)+3}
\end{equation}
We can express the density and pressure in a more convenient way as follows:
\begin{equation}\label{34}
		\rho =\Omega_{m0}\rho_{c0}\frac{H^2 (2 \mu  (q+4)+3)}{H_0^2 (2 \mu  (q_0+4)+3)}
\end{equation}
\begin{equation}\label{35}
	 p = \omega_{m 0}\Omega_{m0}\rho_{c0} \frac{H^2 ((6 \mu +2) q-1)}{H_0^2 ((6 \mu +2) q_0-1)},
\end{equation}
where $\rho_{c0}=\frac{3H_0^2}{8\pi G}$ is the critical density and it has the value $1.88\times 10^{-29}h^2 gm/cm^3$, h=$H_0/100$. The other parameters like $\omega_{m 0}$,  $\Omega_{m0}$, $H_0$ and $ q_0$ have been estimated by us. So we can plot the matter density and pressure. From Eqs. (\ref{34}), (\ref{32}) and (\ref{10}), 
we can plot $\rho_{\mu}$ and $p_{\mu}$. We note that these develop due to the curvature and trace of the  energy momentum tensor dominance in $f(R,T)$ gravity. The followings are the plots:

\begin{figure}[H]	
	(a) \includegraphics[width=8cm,height=7cm,angle=0]{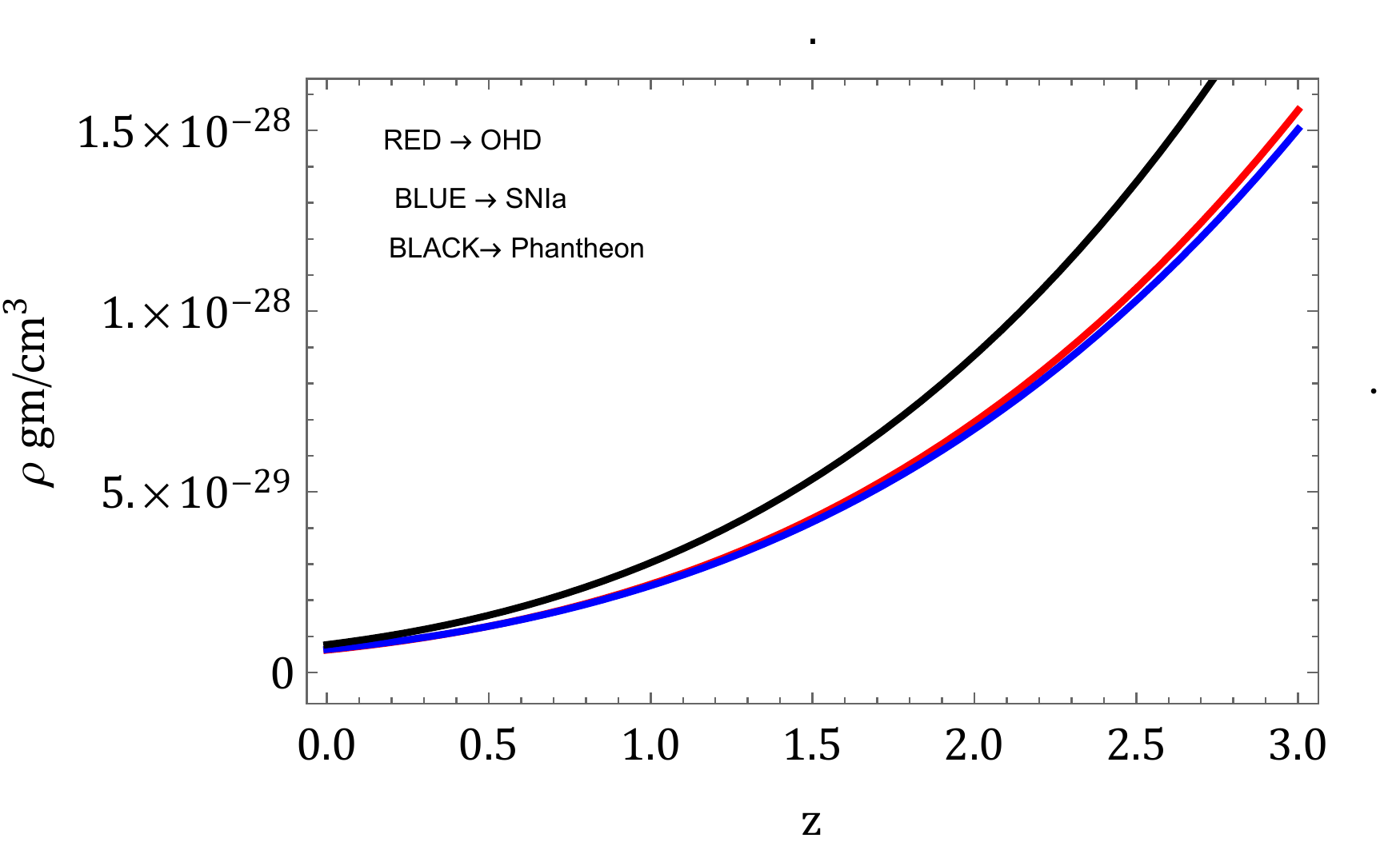}
	(b)\includegraphics[width=8cm,height=7cm,angle=0]{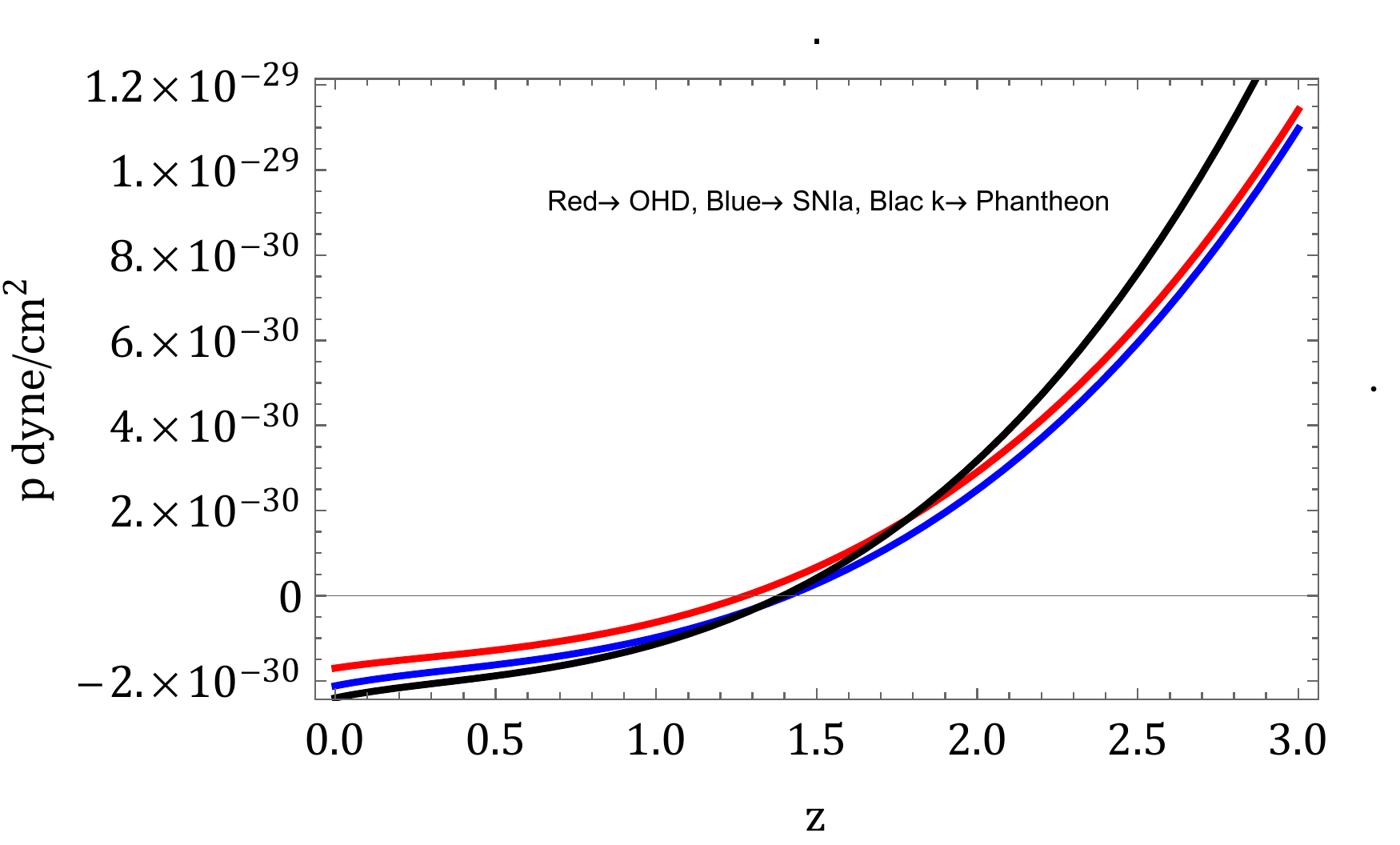}
	(c)\includegraphics[width=8cm,height=7cm,angle=0]{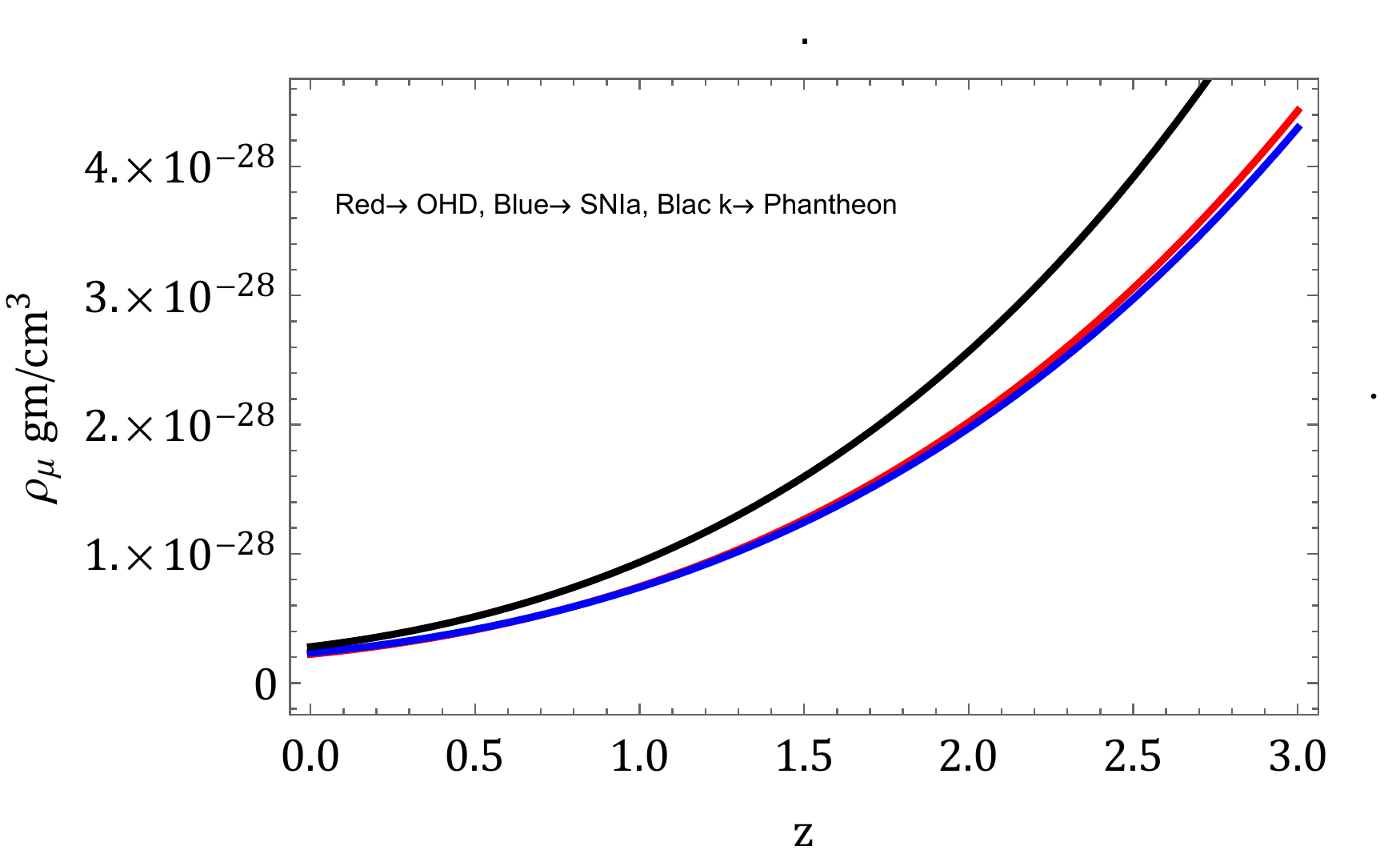}
	(d) \includegraphics[width=8cm,height=7cm,angle=0]{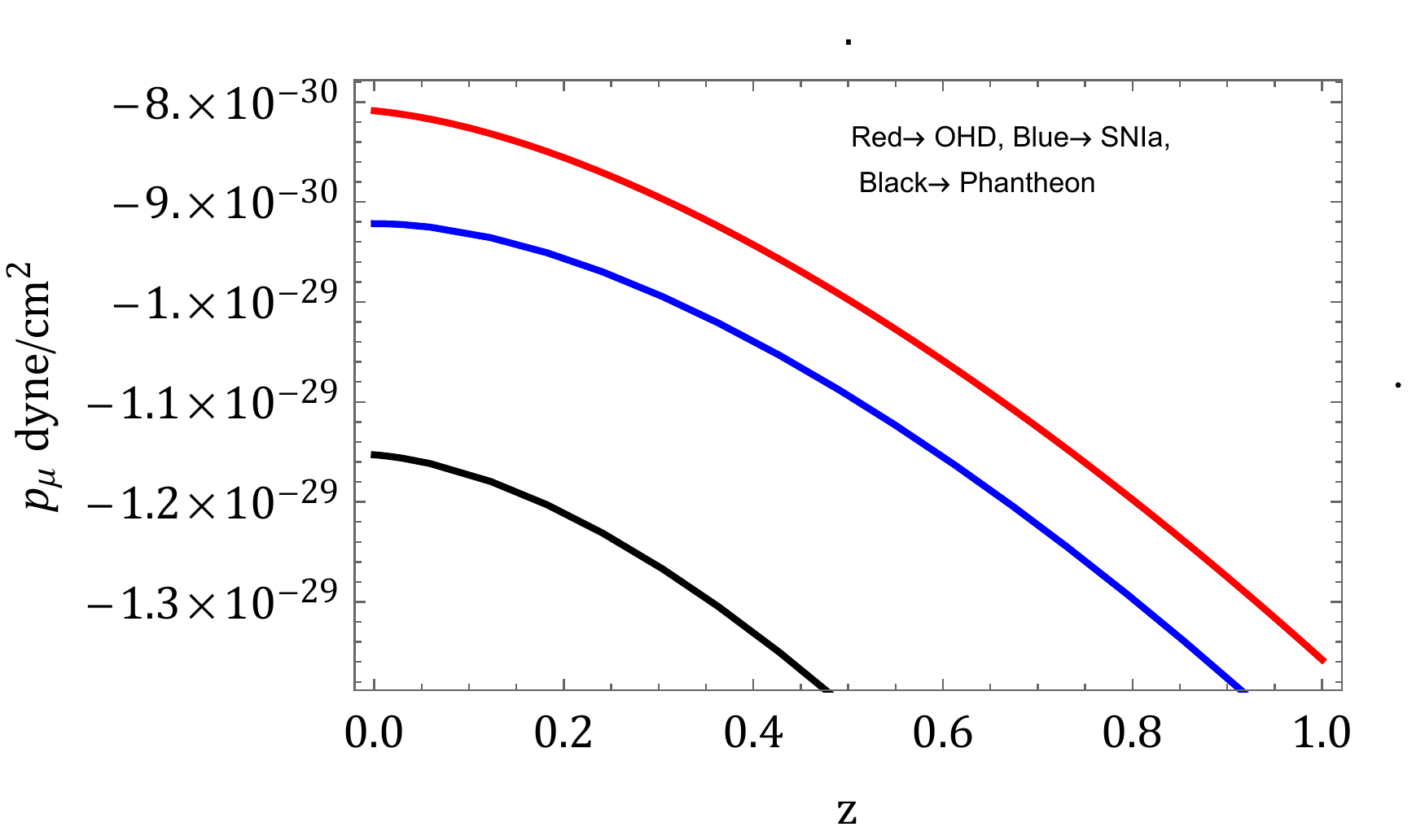}
	\caption{Fig(a) to (d) are plots for the matter and $\mu$ densities and pressures. The $\mu$ pressure is negative, which is responsible for the present day acceleration in the universe. }
\end{figure}

	We make following observations from these the four plots of the Figure $ {\color{red} 9}$.
\begin{itemize}
\item  We can compare the matter and $\mu$ densities in the following way:
$$\rho_{\mu} = \frac{\Omega_{\mu}}{\Omega_m} \rho_m.$$  At present the ratio of the two densities is obtained from Table-1 as $2.22581,~ 2.125~ \text{and}~ 1.60526$ as per the OHD, SNIa and Pantheon data sets, respectively. This shows that quantitatively, the $\mu$ density is more than the matter density.
\item The matter pressure is very small, i.e., dust.
\item  The $\mu$ pressure is negative. This is responsible for the present day acceleration in the universe.
\item Both matter density and pressure are increasing functions of red shift.
\item The $\mu$ density is also showing the same trend as that of the matter.
\end{itemize}

\subsection{Effective density and Effective pressure:} 
 Equation (\ref{12}) may be written as 
 	\begin{equation}\label{36}
 	H^2 (1-2 q)=-8 \pi p_{eff}, ~~ 	3 H^2=8 \pi \rho_{eff} ,
 \end{equation}
where $ p_{eff} = ( p +  p_\mu)$ and $\rho_{eff} =( \rho + \rho_\mu ).$ 
We can  express these in a more convenient way as follows:
\begin{equation}\label{37}
\rho_{eff} = \rho_{c0}	\frac{H^2}{H^2_0}~;~  p_{eff} =  \rho_{c0} 	\frac{H^2(2q-1)}{H^2_0} 
\end{equation}

\begin{figure}[H]	
	(a) \includegraphics[width=8cm,height=7cm,angle=0]{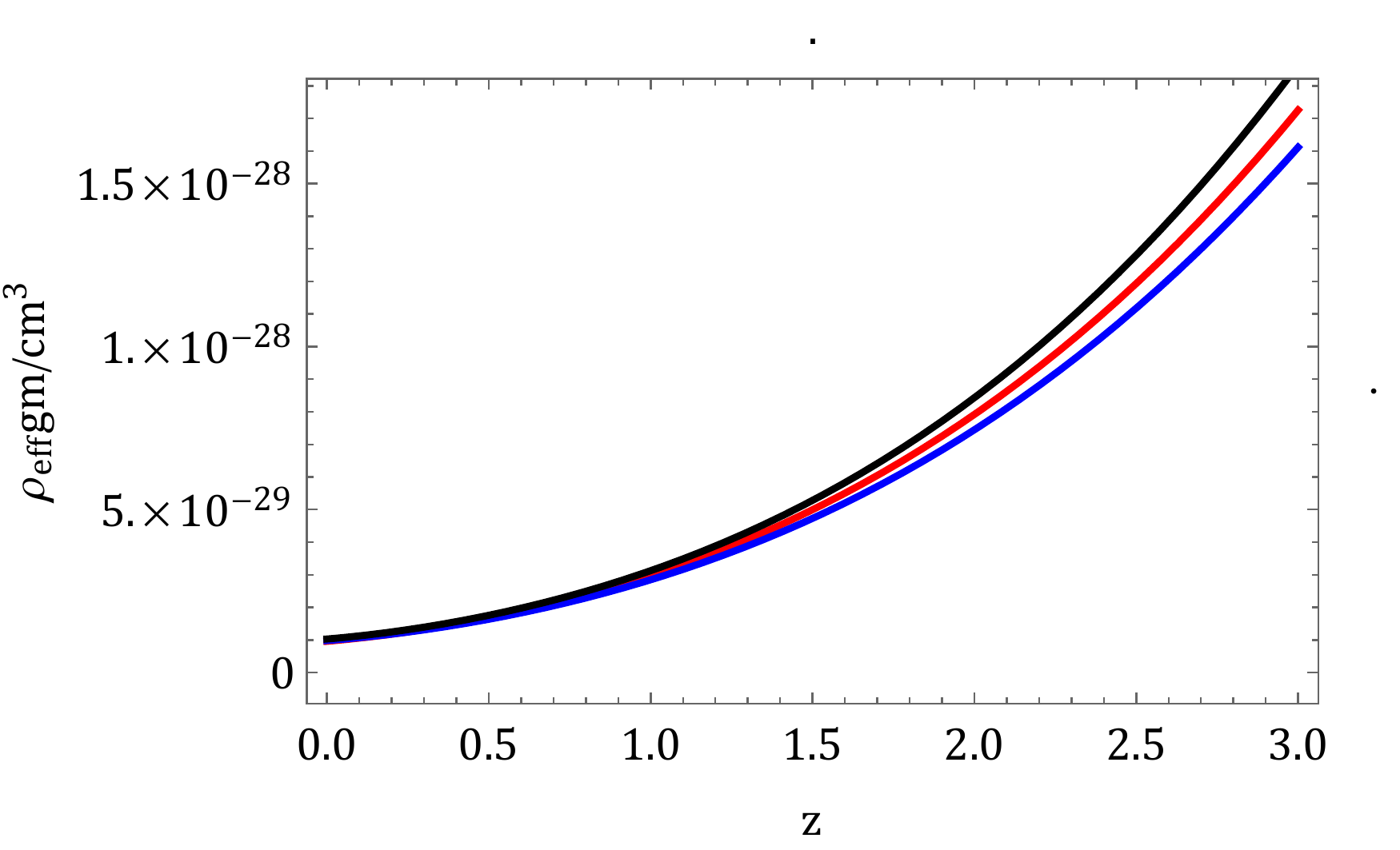}
	(b)\includegraphics[width=8cm,height=7cm,angle=0]{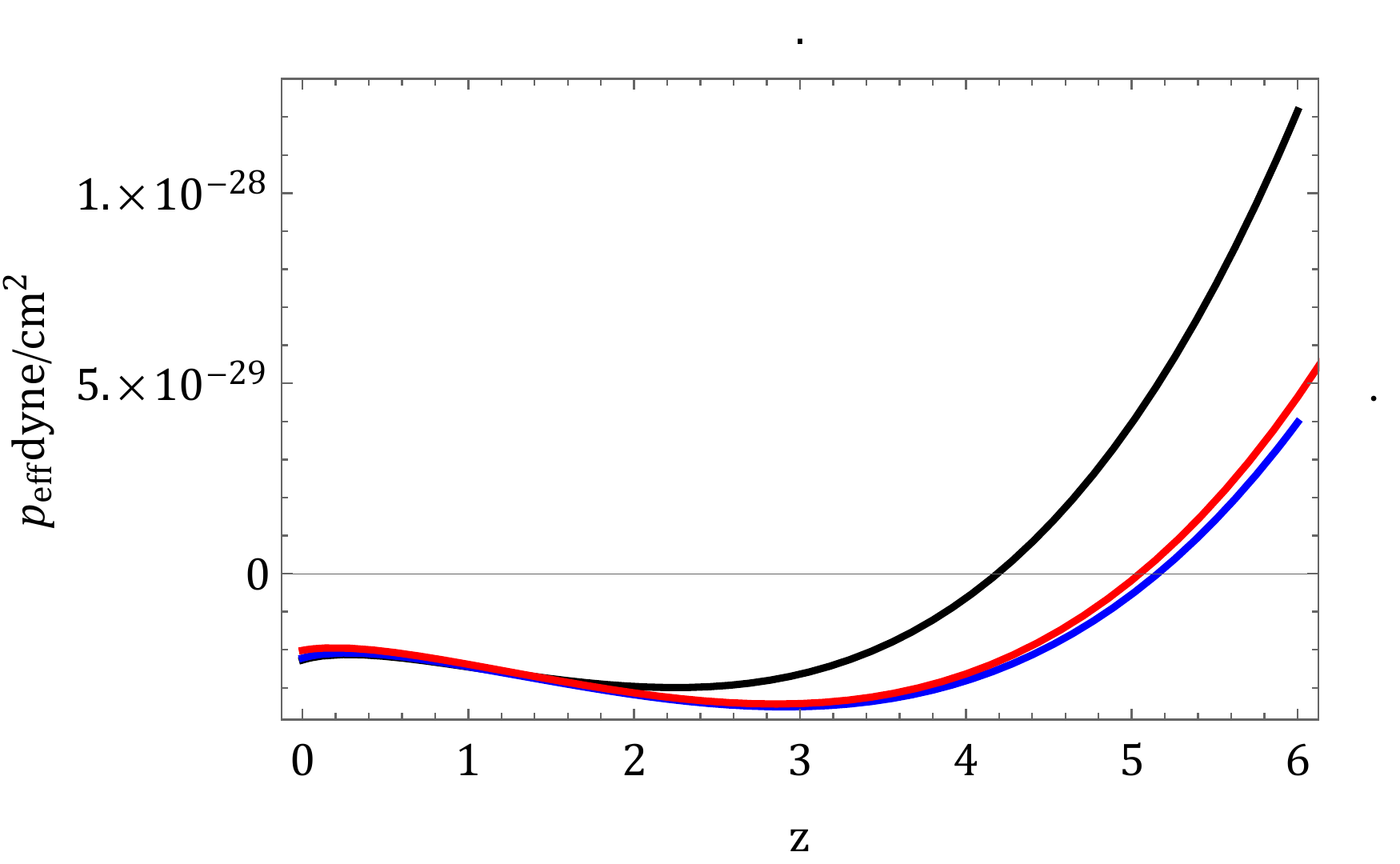}
	(c)\includegraphics[width=8cm,height=7cm,angle=0]{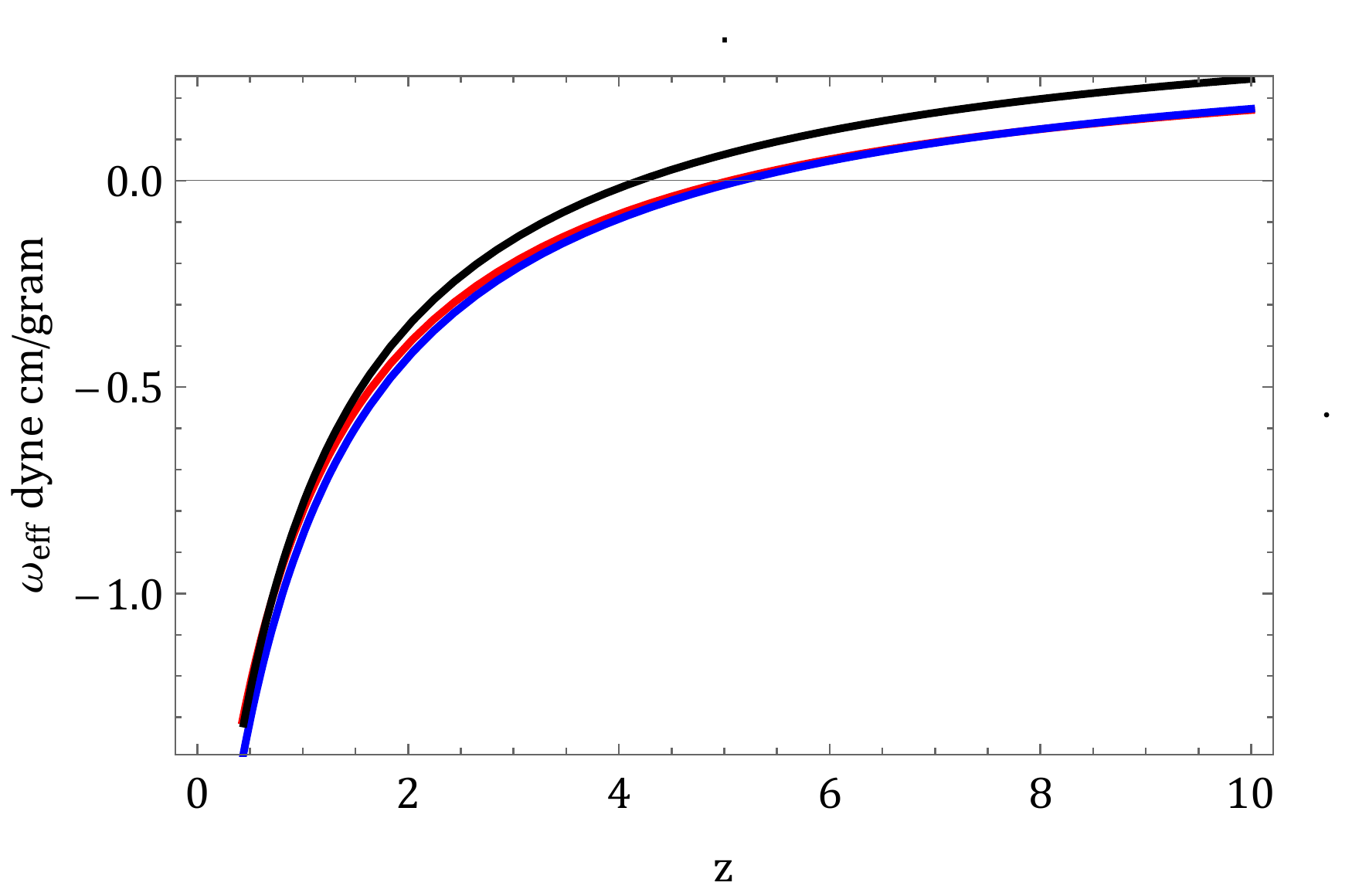}
\caption{Plots (a) to (c) provide an overall picture of our model. Both pressure and equation of state parameter are showing a change from negative to positive, and the  density is increasing with red shift.}
\end{figure}

We make following observations from these the three plots of the Figure $10$:
\begin{itemize}
	\item The density is increasing decreasing with  time. The present value of $\rho_{eff}$ is equal to the critical density. 
	\item  Both pressure and equation of state parameter are showing a change from negative to positive,  showing that our universe is accelerating, and in the past it was decelerating.
\end{itemize}

\subsection{Time versus Red Shift and Transitional times:}
We can calculate the time of any event from the red shift through the following transformation

\begin{equation}\label{38}
	(t_0-t_1)=\intop_{t_1}^{t_0}dt=\intop_{a_1}^{a_0}\frac{da}{aH}=\intop_{0}^{z_1}\frac{dz}{(1+z)H(z)},
\end{equation}
where  $t_0,~ t_1$ is the present time and some past time, respectively. We note that at present, t=$t_0$ and z=0. 
With the help of expression for the Hubble parameter Eq. (\ref{18}), we can plot the graph of time versus red shift. This is given in the following  figure $11$.

\begin{figure}[H]	
	 \includegraphics[width=8cm,height=7cm,angle=0]{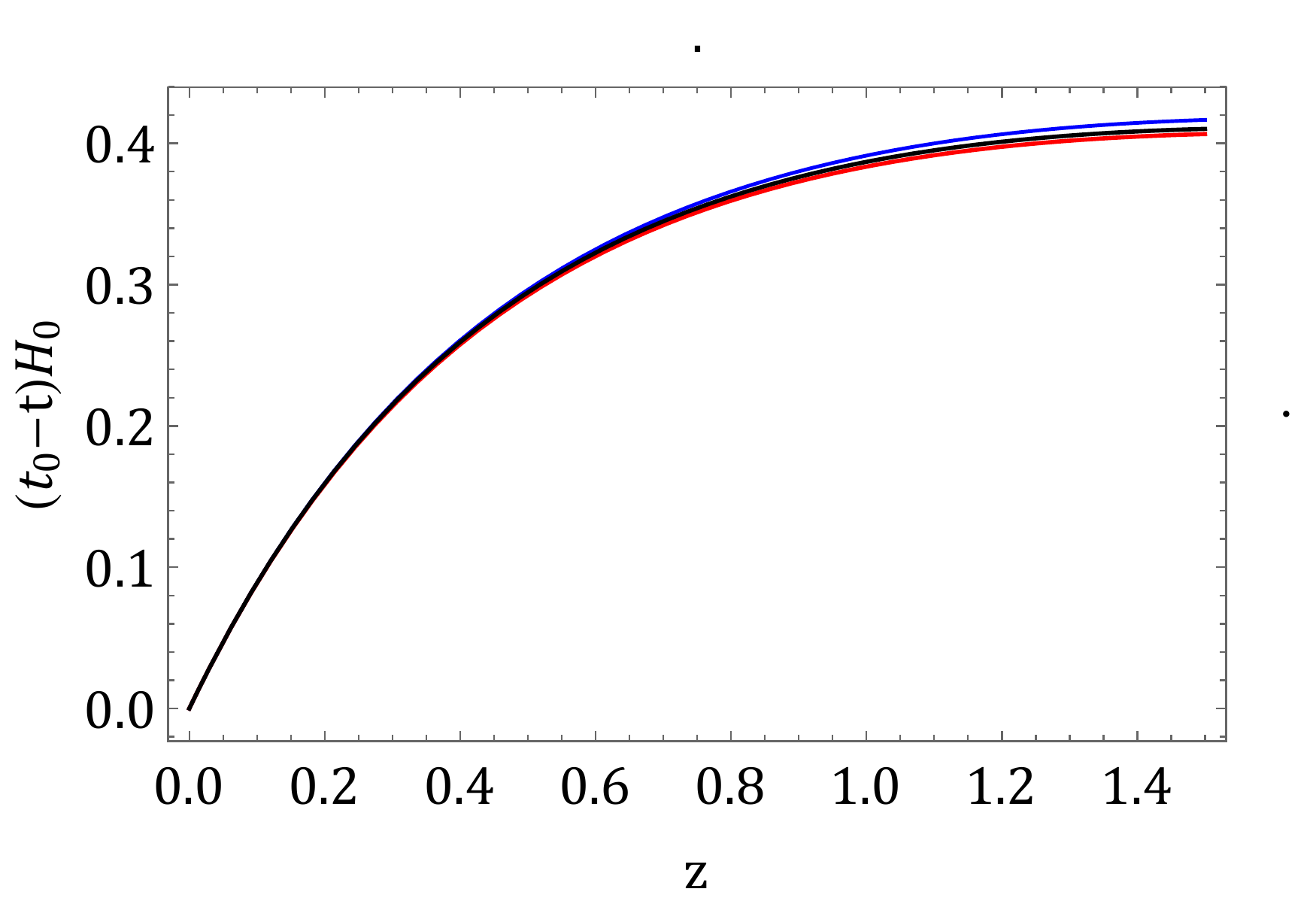}
	
	\caption{The figure describes time versus red shift relation.}
\end{figure}

We make the following observations from  the  plot of the Figure $11$:
\begin{itemize}
	\item The three plots are based on the estimations of the model parameters for the three data sets (i) OHD (ii) SNIa and (iii) pantheon, and they are almost identical.
	\item As $H^{-1}_0 = 9.8 h^{-1}_0\times10^{9}$ yrs, where $h_0=H_0/100$, we can calculate the transition time for the deceleration parameter,  and they are $4.92656\times 10^9$, $4.98347\times10^9$. These correspond to the transition red shifts $z_t= 0.725,~0.741986 ~\text{and}~ 0.8029$.
\end{itemize}
\section{Conclusion:}
        We can summarise our finds as follows:
	    \begin{itemize} 
	 	\item We attempted  to model an FLRW model  filled with a perfect fluid in $f(R,T)$  gravity, where $f(R,T)$ is taken as $R$ + $\frac{\mu}{8\pi}$ T.
	 	\item We  developed two energy parameters $\Omega_m$ and $\Omega_{\mu}$ and used the relation that $\Omega_m$ + $\Omega_{\mu}$=1. The parameter $\Omega_m$ is associated with the matter, whereas $\Omega_{\mu}$ is associated with $f(R,T)$ gravity.  We have statistically estimated that at present $\Omega_{\mu}$ is dominant,  and that the two energy densities are approximately in the ratio 3:1 to 3:2.
	 	\item  We have also introduced $\mu$ pressure ($p_{\mu}$) and $\mu$ density ($\rho_{\mu}$) along with the matter pressure and density. We find that at present the $\mu$ density ($\rho_{\mu}$) is dominant over the matter  density and they are nearly in the ratio  1:3 to 2:3. The negative $\mu$ pressure ($p_{\mu}$) is associated with the acceleration in the universe.
	 	\item  Our model parameters are the present values  of the Hubble, deceleration and equation of state parameters. These parameters were estimated with the help of the three data sets: (i) 77 Hubble OHD data set (ii) 580 SNIa supernova distance modulus data set and (iii)  66 pantheon SNIa which include high red shift data in the range $0\leq z\leq 2.36$.
	 	\item  Our deceleration and snap  parameters show transition from negative to positive and the jerk is always positive. .
	 	\ We have performed a state finder diagnostic of our model  and found  that our model is at present in quintessence.
	 	\item All of our findings are displayed in Table-1 and Table-2. The results are well within the observational findings.
	 	\item All of our outcomes are expressed in terms of red shifts and we have presented transformation to convert red shift into time. Accordingly we have obtained transition times for our deceleration parameter.
	 	 From the cosmological analyses, we infer that modifying Einstein's field equations by replacing  Ricci scalar $R$ by an arbitrary function $f(R,T)$ of  $R$ ans trace $T$ of energy momentum tensor in the Einstein-Hilbert action may lead to arrive at a model which describe transition from deceleration to acceleration at late time. This can be seen as a one step more ahead of the fundamental theory  of general relativity physics on gravitation.
	 	\end{itemize}

        \section*{Acknowledgments}
        The authors (A. Pradhan \& G.K. Goswami ) are grateful for the assistance and facilities provided by the University of Zululand, South Africa during a visit where a part of this article was completed.\\

    \end{document}